\documentclass[12pt,pra,aps,amssymb,amsmath,showpacs,tightenlines]{revtex4}

\newcommand{\half}{\mbox{$\textstyle \frac{1}{2}$}}
\newcommand{\quat}{\mbox{$\textstyle \frac{1}{4}$}}
\newcommand{\octa}{\mbox{$\textstyle \frac{1}{8}$}}
\newcommand{\re}{\mbox{$\rm e$}}
\newcommand{\ri}{\mbox{$\rm i$}}
\newcommand{\rd}{\mbox{$\rm d$}}

\begin{document}

\title{Exactly solvable quantum state reduction models with
time-dependent coupling}

\author{Dorje~C.~Brody${}^*$, Irene~C.~Constantinou${}^*$,
James~D.~C.~Dear${}^\dagger$, and Lane~P.~Hughston${}^\dagger$}

\affiliation{${}^*$Blackett Laboratory, Imperial College, London
SW7 2BZ, UK \\ ${}^\dagger$Department of Mathematics, King's
College London, The Strand, London WC2R 2LS, UK}

\date{\today}

\begin{abstract}
A closed-form solution to the energy-based stochastic
Schr\"odinger equation with a time-dependent coupling is obtained.
The solution is algebraic in character, and is expressed directly
in terms of independent random data. The data consist of (i) a
random variable $H$ which has the distribution ${\mathbb
P}(H=E_i)=\pi_i$, where $\pi_i$ is the transition probability
$|\langle\psi_0|\phi_i\rangle|^2$ from the initial state
$|\psi_0\rangle$ to the L\"uders state $|\phi_i\rangle$ with
energy $E_i$; and (ii) an independent ${\mathbb P}$-Brownian
motion, where ${\mathbb P}$ is the physical probability measure
associated with the dynamics of the reduction process. When the
coupling is time-independent, it is known that state reduction
occurs asymptotically---that is to say, over an infinite time
horizon. In the case of a time-dependent coupling, we show that if
the magnitude of the coupling decreases sufficiently rapidly, then
the energy variance will be reduced under the dynamics, but the
state need not reach an energy eigenstate. This situation
corresponds to the case of a ``partial'' or ``incomplete''
measurement of the energy. We also construct an example of a model
where the opposite situation prevails, in which complete state
reduction is achieved after the passage of a finite period of
time.
\end{abstract}


\maketitle

\section{Introduction}

This paper is concerned with the problem of obtaining closed-form
solutions to the energy-based stochastic extension of the
Schr\"odinger equation in the case of a \emph{time-dependent
coupling parameter}. In this situation the dynamical equation of
the wave function is assumed to satisfy the following stochastic
differential equation:
\begin{eqnarray}
\rd|\psi_t\rangle = -\ri{\hat H}|\psi_t\rangle \rd t - \octa
\sigma_t^2({\hat H}-H_t)^2 |\psi_t\rangle \rd t + \half \sigma_t
({\hat H}-H_t)|\psi_t\rangle \rd W_t. \label{eq:1.1}
\end{eqnarray}
Here $\{|\psi_t\rangle\}_{0\leq t<\infty}$ denotes the
state-vector process, which is defined on a probability space
$(\Omega,{\mathcal F},{\mathbb P})$ with filtration $\{{\mathcal
F}_t\}_{0\leq t<\infty}$, with respect to which $\{W_t\}_{0\leq
t<\infty}$ is a standard one-dimensional Brownian motion. The
operator ${\hat H}$ is the Hamiltonian of the system, and
\begin{eqnarray}
H_t = \langle{\psi}_t|{\hat H}|\psi_t\rangle \label{eq:1.2}
\end{eqnarray}
is the expectation value  of ${\hat H}$ in the state
$|\psi_t\rangle$. The time-dependent coupling parameter
$\{\sigma_t\}_{0\leq t<\infty}$, which has the units $[{\rm
energy}]^{-1}[{\rm time}]^{-1/2}$, and is assumed to be a given
positive function, determines the rate at which state-vector
reduction occurs. For simplicity we consider the case of a
\textit{pure} initial state $|\psi_0\rangle$; the generalisation
to the case of a \textit{mixed} initial state is straightforward.
For convenience we assume that the initial state has norm unity; a
straightforward exercise making use of the Ito calculus then shows
that $\langle\psi_t | \psi_t\rangle=1$ for all $t\geq0$. Likewise
for simplicity we assume that the Hamiltonian ${\hat H}$ has a
discrete spectrum, and that the Hilbert space is of finite
dimension. We assume that the possible values of the energy are
given by $\{E_i\}_{i=1,2,\ldots,N}$. The transition probability
from the initial state $|\psi_0\rangle$ to a state of energy $E_i$
is $\pi_i=\langle\psi_0|{\hat\Pi}_i|\psi_0\rangle$, where
${\hat\Pi}_i$ denotes the projection operator onto the Hilbert
subspace of states with energy $E_i$. Equivalently, we can write
$\pi_i=|\langle\phi_i|\psi_0\rangle|^2$, where $|\phi_i\rangle=
\pi_i^{-1/2}{\hat\Pi}_i|\psi_0\rangle$ is the so-called L\"uders
state associated with the given initial state $|\psi_0\rangle$ and
the energy $E_i$. We note that $\langle\phi_i|\phi_i\rangle=1$.
According to the von Neumann-L\"uders state-vector reduction
hypothesis~\cite{abbh,isham,luders}, if the initial state of the
quantum system is $|\psi_0\rangle$ and a measurement of the energy
is made with the result $E_i$, then there is a discontinuous
transformation of the state of the system, and the new state is
given by $|\phi_i\rangle$. In particular, in the case of a
degenerate spectrum a specific state is selected in this way among
all those with the given eigenvalue.

In the stochastic framework the discontinuous von Neumann-L\"uders
reduction process is replaced by a \textit{continuous} reduction
process modelled by equation (\ref{eq:1.1}). Depending on the
details of the physical setup, the reduction process modelled by
(\ref{eq:1.1}) can be regarded as taking place either (a) as a
consequence of a measurement having been made, or (b) as a result
of interaction of the system with its environment, or (c)
spontaneously. For instance, we can view (\ref{eq:1.1}) as a
phenomenological ``reduced-form'' model for the dynamics of a
system when an energy measurement is made. The time-dependent
coupling in that case represents the exogenous intervention of the
measurement apparatus in the dynamics of the system.

The mathematical and physical properties of the energy-based
stochastic extension of the Schr\"odinger equation have been
studied extensively in the literature \cite{adler,adler2,adler3,
abbh,ah,bh1,bh2,bhs,bh3,bh4,gisin,hughston,pearle1,percival}. The
energy-based stochastic Schr\"odinger equation can be regarded as
a special case of a more general class of stochastic models for
the dynamics of the wave function that use nonlinear stochastic
differential equations of the form (\ref{eq:1.1}), but typically
involving a number of quantum operators. Such models have been
introduced with a variety of aims, and in the general situation
the operators need not commute amongst themselves or with the
Hamiltonian, and need not be Hermitian. We mention, e.g.,
\cite{ab,barchielli,bassi,carmichael2,diosi,gpr,gisin3,pearle,
percival2}, and works cited therein, as references to the
substantial body of publications in this area.

The significance of the energy-based reduction model, in contrast
with the more general situation, is that energy is conserved in
expectation. More specifically, it is a property of the dynamical
equation (\ref{eq:1.1}) that ${\rm tr}({\hat\rho}_t{\hat H})$ is
constant, where the time-dependent density matrix ${\hat\rho}_t$
is defined by ${\hat\rho}_t={\mathbb E}^{{\mathbb
P}}[|\psi_t\rangle \langle\psi_t|]$. Here ${\mathbb E}^{{\mathbb
P}}[-]$ denotes expectation with respect to the physical
probability measure ${\mathbb P}$. Conservation of the energy
follows from the fact that
\begin{eqnarray}
\frac{\rd{\hat\rho}_t}{\rd t} = -\ri [{\hat H},{\hat\rho}_t] +
\quat \sigma_t^2 \left[ {\hat H}{\hat\rho}_t{\hat H} -\half {\hat
H}^2{\hat\rho}_t-\half {\hat\rho}_t{\hat H}^2 \right].
\end{eqnarray}
Thus, the energy-based stochastic Schr\"odinger equation can be
regarded as appropriate to the description of the dynamics of an
isolated system, or any system for which on average there is no
net exchange of energy with the environment.

When the coupling parameter in equation (\ref{eq:1.1}) is a
constant, it is possible to obtain a closed-form solution to this
stochastic differential equation~\cite{bh2,bhs,bh3,bh4}. The
solution to (\ref{eq:1.1}) in the case of a constant coupling
parameter can at each time $t$ be expressed as a function of a
state variable $\xi_t$, the value of which is determined by the
specification of a pair of independent random data corresponding
roughly to the idea of a split between ``signal'' and ``noise''.
More specifically, the state-variable process $\{\xi_t\}_{0\leq
t<\infty}$ is of the form $\xi_t=\sigma H t+B_t$, where $H$ is a
random variable that takes the value $E_i$ $(i=1,2,\ldots,N)$ with
probability $\pi_i$, and $\{B_t\}_{0\leq t<\infty}$ is an
independent Brownian motion. The purpose of this paper is to show
how this construction can be generalised to the case of a generic
time-dependent coupling.

The paper is organised as follows. In Section~\ref{sec:2} we
introduce an ansatz that leads to the solution of (\ref{eq:1.1}).
Some comments are made on the interpretation of the ansatz and its
relation to similar constructions in the theory of filtering. In
Section~\ref{sec:3} we obtain general expressions for the
conditional probability $\pi_{it}={\mathbb E}[{\mathbf 1}_{\{
H=E_i \}}|{\mathcal F}_t]$ and the conditional energy expectation
$H_{t}={\mathbb E}[H|{\mathcal F}_t]$ in terms of the process
$\{\xi_t\}_{0\leq t<\infty}$. Whereas in the time-independent case
the random variable $H_t$ can be expressed, for each value of $t$,
as a function of the state variable $\xi_t$, in the time-dependent
case $H_t$ is a functional of the trajectory $\{\xi_s\}_{0\leq s
\leq t}$. In Section~\ref{sec:4} we show that although $\{\xi_t\}$
is in general non-Markovian, the corresponding energy process
$\{H_t\}$, which is a functional of $\{\xi_t\}$, does have the
Markov property.

In Section~\ref{sec:5} we show how the analysis of
Section~\ref{sec:3} leads to an expression for the state-vector
process that solves (\ref{eq:1.1}). In Section~\ref{sec:65} we
examine the corresponding ``inverse'' problem. We show how, given
any random trajectory $|\psi_t\rangle$ that solves (\ref{eq:1.1}),
it is possible to construct explicitly the independent random data
$H$ and $\{B_t\}$ associated with it. We also remark on the
relation between our solution technique and a well-known
linearisation method often used in analysing the properties of
(\ref{eq:1.1}). Then in Section~\ref{sec:55} we show that the
state vector collapses to the specified eigenstate, provided a
condition on the coupling is satisfied.

More generally, since we consider an essentially arbitrary
time-dependent coupling, one can envisage circumstances in which
the coupling ceases before state reduction is complete. This kind
of situation can be regarded as representing an
\textit{approximate} measurement of the energy, in which partial
information is gained but no definite outcome is obtained. In
Section~\ref{sec:6} we examine this case in some detail, and
derive upper and lower bounds for the asymptotic value of the
energy variance as $t$ goes to infinity.

Another interesting situation arises when the magnitude of the
coupling increases sufficiently rapidly to ensure that state
reduction is completed after the passage of a \textit{finite}
amount of time. In Section~\ref{sec:7} a special example of such a
model is constructed. This example turns out to have a direct
relationship to the finite-time collapse model introduced in
\cite{bh3}. The model presented in \cite{bh3} is based on a
constant coupling parameter and a Brownian bridge noise. Here we
present a alternative model for finite-time collapse, for which
the noise is a standard Brownian motion and the coupling is
time-dependent. We demonstrate that the two models are physically
equivalent.

While the probabilistic method presented in Section~\ref{sec:3} is
effective in obtaining the solution to the problem under
consideration here, there are circumstances in which other methods
are useful as well. We thus outline two further approaches for
obtaining the solution. In Appendix~\ref{app:1} we introduce a
method that involves a discretisation and a continuum limit, and
proceeds in a manner similar to the analysis entailed in the
evaluation of Feynman integrals. The method is computationally
intense, but is satisfying because it allows one to work directly
with the quantities under investigation. In Appendix~\ref{app:2}
we consider another method that is similar to the path integral
approach, except that we use a decomposition of the state-variable
trajectory into increments, and we regard the time-dependent
coupling as moderating the noise rather than the signal.

\section{The quantum information process}
\label{sec:2}

The ansatz that we use to solve (\ref{eq:1.1}) is based on the
specification of a state-variable process $\{\xi_t\}$ which, for
reasons discussed shortly, will be called the ``quantum
information process'', and is of the following form:
\begin{eqnarray}
\xi_t = H \int_0^t \sigma_s \rd s + B_t .  \label{eq:1.3}
\end{eqnarray}
Here $H$ denotes a random variable on the given probability space
$(\Omega,{\mathcal F}, {\mathbb P})$, taking the possible values
$\{E_i\}_{i=1,2,\ldots,N}$ with the probabilities $\{\pi_i\}_{i=
1,2,\ldots,N}$, and $\{B_t\}_{0\leq t<\infty}$ is a standard
Brownian motion, independent of $H$. We do not assume that
$\{B_t\}$ is adapted to the filtration $\{{\mathcal F}_t\}$
introduced earlier. On the contrary, we shall see later that
$\{{\mathcal F}_t\}$ is generated by $\{\xi_t\}$. The various
terms appearing in (\ref{eq:1.3}) can be given an interpretation
in the language of filtering theory. This ``signal and noise"
interpretation, although not essential to the use of the ansatz to
solve (\ref{eq:1.1}), is nonetheless physically very suggestive,
and as a consequence helps to motivate the form that the solution
takes. Indeed, the methodology of filtering theory has been
already shown \cite{bh2} to be effective in deriving solutions to
the energy-based stochastic Schr\"odinger equation, and in what
follows we take this line of investigation further.

The random variable $H$, according to this interpretation,
represents the unknown terminal value of the energy of the quantum
system whose time evolution is described by equation
(\ref{eq:1.1}). The term $H\int_0^t \sigma_s \rd s$ in
(\ref{eq:1.3}) should be thought of as the ``signal'' component of
the quantum information process. As time passes, the magnitude of
the signal component increases, but the true value of $H$ remains
obscured by the presence of a noise process $\{B_t\}_{0\leq t<
\infty}$. The ``accessible'' information concerning the value of
$H$ is represented by the process $\{\xi_t\}_{0\leq t<\infty}$,
which consists of both signal and noise. Given the history
$\{\xi_s\}_{0\leq s<t}$ over a finite time interval $[0,t]$, it is
not generally possible to disentangle the true value of $H$ from
the noise. In the context of filtering theory, the task in such a
setup is to determine the best estimate of $H$, given the
information of the trajectory $\{\xi_s\}_{0\leq s\leq t}$ from
time zero to time $t$. It is a remarkable feature of the
stochastic Schr\"odinger equation that the expectation of the
Hamiltonian turns out to be given by such an estimate.

In the first part of the paper we shall examine the case for which
the state-vector trajectory $\{|\psi_t\rangle\}$ satisfies the
dynamical equation (\ref{eq:1.1}) for all $t\in[0,\infty)$. In order
for the trajectory $\{|\psi_t\rangle\}_{0\leq t<\infty}$ to be
well-defined, the coupling function $\{\sigma_t\}_{0\leq t<\infty}$
must be such that
\begin{eqnarray}
\int_0^t \sigma_s^2 \rd s< \infty, \qquad 0\leq t<\infty.
\label{eq:1.04}
\end{eqnarray}
Additionally, as will be established in Section~\ref{sec:55}, to
ensure a \emph{complete} reduction of the state vector, the
coupling function must be chosen such that
\begin{eqnarray}
\lim_{t\to\infty} \int_0^t \sigma_s^2 \rd s= \infty.
\label{eq:1.4}
\end{eqnarray}
The purpose of this condition is to ensure that the coupling
remains reasonably ``strong'' for all time, and does not attenuate
too much.

There are circumstances in which (\ref{eq:1.04}) is satisfied but
(\ref{eq:1.4}) is not. In such situations the wave function need
not fully collapse to an eigenstate, even though the energy
variance will be reduced. This case is examined in
Section~\ref{sec:6}. It will be assumed throughout Sections
\ref{sec:3}-\ref{sec:6} that $\{\xi_t\}$ and $\{\sigma_t\}$ are
defined for all $t$ in the range $0\leq t<\infty$, and that
(\ref{eq:1.04}) holds. In Section~\ref{sec:7} we drop the
assumption of an infinite collapse time, and consider the case for
which the integral of $\{\sigma_t\}$ diverges after a finite
passage of time.

The estimation problem posed by an ansatz of the form
(\ref{eq:1.3}) is well established in the literature of nonlinear
filtering~\cite{bj,ls}. The relevance of (\ref{eq:1.3}) to the
dynamics of the quantum state $\{|\psi_t\rangle\}$ satisfying the
stochastic Schr\"odinger equation (\ref{eq:1.1}), on the other
hand, is not obvious. As we shall demonstrate, the information
generated by $\{\xi_s\}_{0\leq s\leq t}$ is equivalent to the
information generated by the evolution $\{|\psi_s \rangle\}_{0\leq
s\leq t}$ of the quantum state itself. We formalise this notion by
observing that $\{\xi_t\}_{0\leq t<\infty}$ and
$\{|\psi_t\rangle\}_{0\leq t<\infty}$ generate the same
\emph{filtration} $\{{\mathcal F}_t\}_{0\leq t<\infty}$. As a
consequence, the energy process $\{H_t\}$ determined by the
quantum expectation (\ref{eq:1.2}) of the Hamiltonian operator
turns out to be \emph{indistinguishable} from the process
generated as $t$ varies by the mathematical expectation of the
random variable $H$, conditional on the specification of the
trajectory $\{\xi_s\}_{ 0\leq s\leq t}$.

The best estimate for $H$, in the sense of least quadratic error,
given the history of the information process up to time $t$, is
known (see, e.g., \cite{bh2}) to be the conditional expectation
\begin{eqnarray}
H_t = {\mathbb E}\left[H|\{\xi_s\}_{0\leq s\leq t}\right].
\label{eq:1.5}
\end{eqnarray}
We have used the same notation $\{H_t\}$ for the processes defined
in (\ref{eq:1.2}) and in (\ref{eq:1.5}) because these processes
will be shown to be the same. When $\{\sigma_t\}$ is constant,
equation (\ref{eq:1.5}) can be simplified to the form
$H_t={\mathbb E}\left[H|\xi_t\right]$. In this case, $\{\xi_t\}$
is Markovian: this is the situation considered in
\cite{bh2,bhs,bh3, bh4}. However, if $\{\sigma_t\}$ is not
constant, then in general the trajectory $\{\xi_s\}_{0\leq s\leq
t}$ must be taken into account to determine the conditional
expectation (\ref{eq:1.5}). The non-Markovian property of
$\{\xi_t\}$ can be seen intuitively as follows. Writing
(\ref{eq:1.3}) in differential form, we have
\begin{eqnarray}
\rd\xi_t = \sigma_t H \rd t+\rd B_t, \label{eq:1.55}
\end{eqnarray}
which makes it evident that $\{\sigma_t\}$ determines the
\textit{strength} of the signal, that is to say, the rate at which
the true value of $H$ is revealed. If $\{ \sigma_t\}$ is constant,
then sampling from $\{\xi_t\}$ at any small time period in the
interval $[0,t]$ is as good as any other. This is, in essence, the
property of $\{\xi_t\}$, when $\{\sigma_t\}$ is constant, that
makes it Markovian. If $\{\sigma_t\}$ is not constant, then there
is a temporal bias in the sampling from $\{\xi_t\}$, and
observations from different periods cannot be treated on an equal
footing.

\section{Conditional probability process}
\label{sec:3}

Our goal in this section is to work out an explicit expression for
the conditional probability process $\{\pi_{it}\}$ defined by
\begin{eqnarray}
\pi_{it}={\mathbb E}\left[ {\mathbf 1}_{\{H=E_i\}}|
\{\xi_s\}_{0\leq s\leq t}\right]. \label{eq:3.0}
\end{eqnarray}
Here ${\mathbf 1}_{\{A\}}$ denotes the indicator function which
takes the value unity if $A$ is true, and zero if $A$ is false.
Once we obtain $\pi_{it}$, then the conditional expectation
(\ref{eq:1.5}) can be obtained by the relation $H_t=\sum_i
\pi_{it}E_i$.

To determine $\{\pi_{it}\}$ we use a change-of-measure technique
(see, e.g., \cite{bj,davis,kailath}), proceeding as follows. Let
$(\Omega,{\mathcal F}, {\mathbb P})$ be a probability space on
which a standard Brownian motion $\{B_t\}$ is defined, and let $H$
be a random variable on $(\Omega,{\mathcal F}, {\mathbb P})$ that
is independent of $\{B_t\}$. We fix a time interval $[0,u]$, and
let $\{\xi_t\}_{0\leq t\leq u}$ be given by
\begin{eqnarray}
\xi_t = H\int_0^t \sigma_s\rd s + B_t, \label{eq:3.1}
\end{eqnarray}
where $\{\sigma_t\}$ is deterministic and satisfies
(\ref{eq:1.04}). Next, we define a process $\{\Lambda_t\}_{0 \leq
t\leq u}$ over $[0,u]$ by the expression
\begin{eqnarray}
\Lambda_t = \exp\left( H\int_0^t\sigma_s\rd \xi_s-\half H^2
\int_0^t \sigma_s^2\rd s\right), \label{eq:3.5}
\end{eqnarray}
or equivalently, by virtue of (\ref{eq:3.1}),
\begin{eqnarray}
\Lambda_t^{-1} = \exp\left( -H\int_0^t\sigma_s\rd B_s-\half H^2
\int_0^t \sigma_s^2\rd s\right). \label{eq:3.2}
\end{eqnarray}

The idea is to use $\{\Lambda_t\}$ to make a change of probability
measure. The new probability measure will be defined on the space
$(\Omega,{\mathcal G}_{u})$, where ${\mathcal G}_{u}\subset
{\mathcal F}$ is the $\sigma$-subalgebra of events determined by
the specification of the trajectory $\{B_t\}_{0 \leq t\leq u}$
over the given time horizon, together with $H$.

We recall that the points of $\Omega$ represent the possible
outcomes of chance, and the elements of ${\mathcal F}$ are subsets
of $\Omega$ with the property that for each such subset $A\in
{\mathcal F}$ the measure ${\mathbb P}$ assigns a probability
${\mathbb P}(A)$ to the event that $\omega\in A$. The elements of
${\mathcal G}_{u}$ consist of those elements $A\in {\mathcal F}$
with the property that knowledge of the value of $H$ and the
trajectory $\{B_t\}_{0 \leq t\leq u}$ is sufficient to determine
whether $\omega\in A$. For any element $A\in {\mathcal G}_{u}$ the
value of the indicator function ${\mathbf 1}_{\{\omega\in A\}}$ is
determined by the specification of $H$ and $\{B_t\}_{0 \leq t\leq
u}$. The new measure ${\mathbb Q}$ on $(\Omega,{\mathcal G}_{u})$
is then given as follows. For any set $A\in {\mathcal G}_{u}$ we
define
\begin{eqnarray}
{\mathbb Q}(A)={\mathbb E}^{\mathbb P}\left[ \Lambda_{u}^{-1}
{\mathbf 1}_{\{\omega\in A\}}\right] .
\end{eqnarray}
This relation is usually abbreviated by writing $\rd{\mathbb Q}=
\Lambda_{u}^{-1}\rd{\mathbb P}$. Since ${\mathbb E}^{\mathbb P}
\left[\Lambda_{u}^{-1}\right]=1$ by virtue of elementary
properties of the stochastic exponential (\ref{eq:3.2}), it
follows that ${\mathbb Q}(\Omega)=1$.

Given the setup described above, we have the following facts: (i)
on the probability space $(\Omega,{\mathcal G}_{u}, {\mathbb Q})$,
the process $\{\xi_t\}_{0\leq t\leq u}$ defined by (\ref{eq:3.1})
is a Brownian motion, and is independent of $H$; (ii) the random
variable $H$ has the same probability law with respect to
${\mathbb Q}$ as it does with respect to ${\mathbb P}$; (iii) for
all $t\in[0,u]$, the conditional expectation $f_t={\mathbb
E}^{\mathbb P}[f(H)|\{\xi_s\}_{0\leq s\leq t}]$ of a function of
the random variable $H$ can be expressed in the form
\begin{eqnarray}
f_t = \frac{{\mathbb E}^{\mathbb Q}\left[f(H)\Lambda_t|
\{\xi_s\}_{0\leq s\leq t}\right]}{{\mathbb E}^{\mathbb Q}
\left[\Lambda_t|\{\xi_s\}_{0\leq s\leq t}\right]} .\label{eq:3.4}
\end{eqnarray}

To work out an expression for $\pi_{it}$ we start with the
definition (\ref{eq:3.0}), substitute (\ref{eq:3.5}) into
(\ref{eq:3.4}) and set $f(H)={\mathbf 1}_{\{H=E_i\}}$ to obtain
\begin{eqnarray}
\pi_{it}= \frac{ \pi_i \exp\left(E_i\int_{0}^{t}\sigma_{s}\rd
\xi_{s} -\half E_i^2 \int_{0}^{t}\sigma_{s}^2\rd s\right)} {\sum_i
\pi_i \exp\left(E_i\int_{0}^{t}\sigma_{s}\rd \xi_{s} -\half E_i^2
\int_{0}^{t}\sigma_{s}^2\rd s\right)}.  \label{eq:2.26a}
\end{eqnarray}
Similarly, by setting $f(H)=H$ we obtain
\begin{eqnarray}
H_t= \frac{\sum_i \pi_i E_i \exp\left(E_i\int_{0}^{t}\sigma_{s}\rd
\xi_{s} -\half E_i^2 \int_{0}^{t}\sigma_{s}^2\rd s\right)} {\sum_i
\pi_i \exp\left(E_i\int_{0}^{t}\sigma_{s}\rd \xi_{s} -\half E_i^2
\int_{0}^{t}\sigma_{s}^2\rd s\right)}. \label{eq:3.7}
\end{eqnarray}
The result appears at first glance to depend on the choice of time
horizon $u$, since $\{\pi_{it}\}$ and $\{H_t\}$ are only defined
for $t\in[0,u]$; but it is straightforward to see that
(\ref{eq:2.26a}) and (\ref{eq:3.7}) remain valid for all $t\in[0,
\infty)$.

The arguments establishing the validity of (i), (ii), and (iii)
above can be sketched briefly as follows. First we note that the
relation between expectation under the ${\mathbb P}$-measure and
expectation under the ${\mathbb Q}$-measure is given by
\begin{eqnarray}
{\mathbb E}^{\mathbb Q}[X] = {\mathbb E}^{\mathbb P}[
\Lambda^{-1}_u X] . \label{eq:sup1}
\end{eqnarray}
To see that $\{\xi_t\}_{0\leq t<\infty}$ is a Brownian motion on
$(\Omega, {\mathcal F},{\mathbb Q})$ it suffices to show that
\begin{eqnarray}
{\mathbb E}^{\mathbb Q}\left[\re^{x\xi_s+y\xi_t}\right] =
\re^{\frac{1}{2} \left(x^2s+y^2t+2xys \right)} \label{eq:sup2}
\end{eqnarray}
for $s\leq t$ and for all $x,y$. In particular, if $\{\xi_t\}$
possesses this bi-characteristic function, then it follows at once
that $\{\xi_t\}$ is ${\mathbb Q}$-Gaussian, and that ${\rm Cov}
[\xi_s,\xi_t]=s$ for $s\leq t$. These properties, together with
the fact that $\{\xi_t\}$ is continuous, are sufficient to
characterise it as a Brownian motion under ${\mathbb Q}$. The
verification of (\ref{eq:sup2}) follows by a calculation that
makes use of (\ref{eq:3.1}), (\ref{eq:3.2}), and (\ref{eq:sup1}),
the ${\mathbb P}$-independence of $H$ and $\{B_t\}$, and basic
properties of $\{B_t\}$ under ${\mathbb P}$. The ${\mathbb
Q}$-independence of $H$ and $\{\xi_t\}$ then follows by a similar
calculation that establishes that
\begin{eqnarray}
{\mathbb E}^{\mathbb Q}\left[\re^{x\xi_t+y H}\right] =
\re^{\frac{1}{2}x^2t} \sum_i \pi_i \, \re^{y E_i}. \label{eq:sup3}
\end{eqnarray}
To show that the probability law of $H$ is the same under
${\mathbb P}$ and ${\mathbb Q}$ we make use of the ${\mathbb
P}$-independence of $H$ and $\{B_t\}$ to observe that
\begin{eqnarray}
{\mathbb Q}\left(H=E_i\right) &=& {\mathbb E}^{\mathbb P} \left[
\Lambda_u^{-1} {\bf 1}_{\{H=E_i\}} \right] \nonumber \\ &=&
{\mathbb E}^{\mathbb P} \left[ {\bf 1}_{\{H=E_i\}} \exp\left(
-H\int_0^t\sigma_s\rd B_s-\half H^2 \int_0^t \sigma_s^2\rd
s\right) \right] \nonumber \\ &=& {\mathbb E}^{\mathbb P} \left[
{\bf 1}_{\{H=E_i\}} \exp\left( -E_i\int_0^t\sigma_s\rd B_s-\half
E_i^2 \int_0^t \sigma_s^2\rd s\right) \right] \nonumber \\ &=&
{\mathbb E}^{\mathbb P} \left[ {\bf 1}_{\{H=E_i\}} \right]
{\mathbb E}^{\mathbb P} \left[ \exp\left( -E_i\int_0^t\sigma_s\rd
B_s-\half E_i^2 \int_0^t \sigma_s^2\rd s\right) \right] \nonumber \\
&=&  {\mathbb P}\left(H=E_i\right). \label{eq:sup4}
\end{eqnarray}

As for (iii), we remark first that relation (\ref{eq:3.4}) is a
special case of a more general result referred to as the
Kallianpur-Striebel formula~\cite{ks}. A derivation of
(\ref{eq:3.4}) can be sketched as follows. We reverse the
construction above and start with a probability space
$(\Omega,{\mathcal F}, {\mathbb Q})$ on which $\{\xi_t\}_{0 \leq
t<\infty}$ is a standard Brownian motion, and $H$ an independent
random variable taking the values $\{E_i\}_{i=1,2,\ldots,N}$ with
the probabilities $\{\pi_i\}_{i=1,2,\ldots,N}$. We fix a time
interval $[0,u]$ and let ${\mathcal G}_{u}$ denote the
$\sigma$-subalgebra of events generated by $\{\xi_t\}_{0 \leq
t\leq u}$ and $H$. Assuming that $\{\sigma_t\}$ satisfies
(\ref{eq:1.04}), we define the process $\{\Lambda_t\}_{0 \leq
t\leq u}$ by (\ref{eq:3.5}), as before, and we introduce the
measure ${\mathbb P}$ by setting $\rd{\mathbb P}=\Lambda_{u}
\rd{\mathbb Q}$, or equivalently, ${\mathbb P}(A)={\mathbb
E}^{\mathbb Q}\left[ \Lambda_{u} {\mathbf 1}_{\{\omega\in
A\}}\right]$. Then the process $\{B_t\}_{0 \leq t\leq u}$ defined
by $B_t=\xi_t-H\int_0^t\sigma_s\rd s$ is a Brownian motion with
respect to ${\mathbb P}$, and is ${\mathbb P}$-independent of $H$.
It is worthwhile emphasising that ${\mathcal F}_t=\sigma\left(
\{\xi_s\}_{0\leq s\leq t}\right)$, whereas ${\mathcal G}_t=\sigma
\left( H,\{\xi_s\}_{0\leq s\leq t}\right) = \sigma \left( H,\{B_s
\}_{0\leq s\leq t}\right)$. Thus ${\mathcal F}_t\subset{\mathcal
G}_t$. One can think of $\{{\mathcal F}_t\}$ as the filtration
generated by the dynamics of the state vector process
$\{|\psi_t\rangle\}_{0 \leq t<\infty}$; whereas in the larger
``nonphysical'' filtration $\{{\mathcal G}_t\}$ the value of $H$
is already ``known'' at time $0$. It follows by the conditional
form of the change of measure relation that
\begin{eqnarray}
{\mathbb E}^{\mathbb P}\left[f(H)|\{\xi_s\}_{0\leq s\leq t}\right] =
\frac{{\mathbb E}^{\mathbb Q}\left[f(H)\Lambda_{u}| \{\xi_s\}_{0\leq
s\leq t}\right]}{{\mathbb E}^{\mathbb Q}
\left[\Lambda_{u}|\{\xi_s\}_{0\leq s\leq t}\right]}. \label{eq:3.8}
\end{eqnarray}
Finally, we use the fact that $H$ and $\{\xi_s\}_{0\leq s\leq u}$
are ${\mathbb Q}$-independent to deduce that
\begin{eqnarray}
{\mathbb E}^{\mathbb Q} \left[f(H)\Lambda_{u}|\{\xi_s\}_{0\leq
s\leq t}\right]={\mathbb E}^{\mathbb Q} \left[f(H)\Lambda_t|
\{\xi_s\}_{0\leq s\leq t} \right] \label{eq:3.9}
\end{eqnarray}
for any choice of $f(H)$, and as a consequence we deduce
(\ref{eq:3.4}).

\section{On the Markovian nature of the energy process}
\label{sec:4}

Before we verify that our expression (\ref{eq:3.7}) for the
conditional expectation of $H$ agrees with the energy process
(\ref{eq:1.2}), it will be useful to show that $\{H_t\}$, as
defined by (\ref{eq:3.7}), has the Markov property. We note in
particular that by virtue of (\ref{eq:3.7}) we have
\begin{eqnarray}
H_t= \frac{\sum_i \pi_i E_i \exp\left(E_i\eta_{t}-\half E_i^2
\int_{0}^{t}\sigma_{s}^2\rd s\right)} {\sum_i \pi_i
\exp\left(E_i\eta_{t} -\half E_i^2 \int_{0}^{t}\sigma_{s}^2\rd
s\right)}, \label{eq:3.75}
\end{eqnarray}
where the process $\{\eta_t\}_{0\leq t<\infty}$ is defined by
\begin{eqnarray}
\eta_t = \int_0^t\sigma_s\rd \xi_s .\label{eq:3.6}
\end{eqnarray}
To show that $\{H_t\}$ is a Markov process it will suffice if we
can show: (i) that $\{\eta_t\}$ is a Markov process; and (ii) that
$\eta_t$ can be expressed as a function of $H_t$, i.e. that the
relation between $\eta_t$ and $H_t$ is invertible. In particular,
if the relation between $\eta_t$ and $H$ is invertible, then the
filtration generated by $\{\eta_t\}$ is the same as the filtration
generated by $\{H_t\}$.

To verify that $\{\eta_t\}$ is Markovian we must demonstrate that,
for all $T\geq t$ and for all $x\in{\mathbb R}$, we have
\begin{eqnarray}
{\mathbb P}(\eta_T\leq x|\{\eta_s\}_{0\leq s\leq t})={\mathbb P}
(\eta_T \leq x|\eta_t). \label{eq:4.2}
\end{eqnarray}
Alternatively, it suffices to show that
\begin{eqnarray}
{\mathbb P}\left(\eta_t\leq x|\eta_s,\eta_{s_1},\eta_{s_2},
\ldots, \eta_{s_k}\right) = {\mathbb P}\left( \eta_t\leq x|\eta_s
\right) \label{eq:4.3}
\end{eqnarray}
for any collection of times $t,s,s_1,s_2,\ldots,s_k$ such that
$t\geq s\geq s_1\geq s_2\geq\cdots\geq s_k>0$. To check that these
conditions are satisfied we proceed as follows. First we note that
by virtue of (\ref{eq:3.6}) we have
\begin{eqnarray}
\eta_t = H \int_0^t \sigma_s^2\rd s + \int_0^t\sigma_s\rd B_s,
\label{eq:4.1}
\end{eqnarray}
and therefore
\begin{eqnarray}
\frac{\eta_u}{\int_0^u \sigma_s^2\rd s} - \frac{\eta_v}{\int_0^v
\sigma_s^2\rd s} = \frac{\int_0^u \sigma_s\rd B_s}{\int_0^u
\sigma_s^2\rd s} - \frac{\int_0^v \sigma_s\rd B_s}{\int_0^v
\sigma_s^2\rd s} \label{eq:4.1y}
\end{eqnarray}
for all $u\geq v>0$. We shall establish that the process
$\{\varphi_u\}_{u>0}$ defined by
\begin{eqnarray}
\varphi_u = \frac{\int_0^u \sigma_s\rd B_s}{\int_0^u \sigma_s^2\rd
s} \label{eq:4.4}
\end{eqnarray}
appearing in (\ref{eq:4.1y}) has independent increments. In
particular, since $\{\varphi_u\}$ is a Gaussian process it
suffices to show that $\varphi_b-\varphi_a$ and
$\varphi_d-\varphi_c$ are independent for all $d\geq c\geq b\geq
a>0$. To see this, we note that since $\varphi_b-\varphi_a$ and
$\varphi_d-\varphi_c$ are Gaussian random variables, for their
independence it is sufficient to verify that the covariance
${\mathbb E} [(\varphi_b-\varphi_a) (\varphi_d-\varphi_c)]$
vanishes. But this follows after a short calculation making use of
the Wiener-Ito isometry
\begin{eqnarray}
{\mathbb E}\left[\left(\int_0^u\sigma_s\rd B_s\right) \left(
\int_0^v\sigma_s\rd B_s\right)\right]=\int_0^{u\wedge v}\!
\sigma_s^2 \rd s, \label{eq:4.6}
\end{eqnarray}
where $u\wedge v=\min(u,v)$. Similarly, one can verify that
$\varphi_c$ is independent of $\varphi_b-\varphi_a$ for all $c\geq
b\geq a>0$. This follows from the fact that the increment
$\varphi_\infty-\varphi_c$ is independent of
$\varphi_b-\varphi_a$. Next we observe that as a consequence of
the definitions of $\{\eta_t\}$ and $\{\varphi_t\}$ we have
\begin{eqnarray}
{\mathbb P}\left( \eta_t\leq x| \eta_s,\eta_{s_1}, \eta_{s_2},
\ldots,\eta_{s_k}\right)={\mathbb P}\left(\eta_t\leq x\left| \eta_s,
\varphi_s-\varphi_{s_1}, \varphi_{s_1}- \varphi_{s_2}, \ldots,
\varphi_{s_{k-1}} - \varphi_{s_k}\right.\right). \label{eq:4.7}
\end{eqnarray}
However, since $\eta_t$ and $\eta_s$ are independent of
$\varphi_s- \varphi_{s_1}$, $\varphi_{s_1}-\varphi_{s_2}$,
$\ldots$, $\varphi_{s_{k-1}}-\varphi_{s_k}$, the desired result
(\ref{eq:4.3}) follows.

To show that $H_t$ is invertible as a function of $\eta_t$ it will
suffice to show that for each fixed $t$ the function
\begin{eqnarray}
H(\eta,t) = \frac{\sum_i \pi_i E_i \exp\left(E_i\eta-\half E_i^2
\int_{0}^{t}\sigma_{s}^2\rd s\right)} {\sum_i \pi_i
\exp\left(E_i\eta -\half E_i^2 \int_{0}^{t}\sigma_{s}^2\rd
s\right)} \label{eq:3.755}
\end{eqnarray}
is monotonic in $\eta$. But this can be seen immediately, since
\begin{eqnarray}
\frac{\partial H(\eta,t)}{\partial \eta} = \frac{\sum_i \pi_i (E_i
- H(\eta,t))^2 \exp\left(E_i\eta-\half E_i^2
\int_{0}^{t}\sigma_{s}^2\rd s\right)} {\sum_i \pi_i
\exp\left(E_i\eta -\half E_i^2 \int_{0}^{t}\sigma_{s}^2\rd
s\right)}, \label{eq:3.7555}
\end{eqnarray}
which is positive for all values of $\eta$. As a consequence we
see that the process $\{H_t\}$ as defined by (\ref{eq:3.7}) has
the Markov property. In particular, we have the following
equalities: ${\mathbb P}\left(H_T\leq x|\{H_s\}_{0\leq s\leq t}
\right) = {\mathbb P}\left(H_T\leq x|H_t \right) = {\mathbb P}
\left(H_T\leq x|\eta_t \right) = {\mathbb P} \left(H_T\leq x|
{\mathcal F}_t \right)$.

It is interesting to note that the conditional probability process
$\{\pi_{it}\}$ given by (\ref{eq:2.26a}) can be expressed in the
form
\begin{eqnarray}
\pi_{it}= \frac{\pi_{is}\exp\left(E_i\int_{s}^{t}\sigma_{u}\rd \xi_u
-\half E_i^2 \int_{s}^{t}\sigma_{u}^2\rd u\right)} {\sum_i \pi_{is}
\exp\left(E_i\int_{s}^{t}\sigma_{u}\rd \xi_{u} -\half E_i^2
\int_{s}^{t}\sigma_{u}^2\rd u\right)}. \label{eq:4.8}
\end{eqnarray}
The interpretation of this relation is that the energy-based
reduction models exhibit a fundamental \emph{dynamic consistency}
property. In other words, if at some intermediate time $s$, where
$0<s<t$, we take note of the \textit{a posteriori} conditional
probability $\pi_{is}$, which is based on all information
available up to time $s$, then we see that the resulting ``new''
model for the collapse of the wave function, given by
(\ref{eq:4.8}), is of exactly the same form as the original model,
with $\pi_{is}$ playing the role of the new \textit{a priori}
probability. This means that the choice of the initial time $t =
0$ has no preferential status in the theory. Indeed, the dynamical
consistency of the energy-based reduction theory shows that the
objections raised by Pearle \cite{pearle1} in this connection are
groundless. It is perfectly consistent to regard the collapse
process as having already started at some earlier time than ``the
present''.

\section{Innovation process and solution}
\label{sec:5}

In Section~\ref{sec:3} we calculated the expectation of the random
variable $H$ conditional on the specification of $\{\xi_s
\}_{0\leq s\leq t}$, and we claimed that the result gives the
energy expectation process (\ref{eq:1.2}). The aim of this section
is to verify this claim, by showing how $\{\xi_t\}$ is related to
the Brownian motion $\{W_t\}$ of equation (\ref{eq:1.1}). We begin
by analysing the dynamics of the energy process (\ref{eq:3.7}). A
direct application of Ito's rule shows that
\begin{eqnarray}
\rd H_t=-\sigma_t^2 H_t V_t \rd t + \sigma_t V_t \rd\xi_t,
\label{eq:5.1}
\end{eqnarray}
where
\begin{eqnarray}
V_t = {\mathbb E}\left[(H-H_t)^2|\{\xi_s\}_{0\leq s\leq t}\right]
\label{eq:5.2}
\end{eqnarray}
is the conditional variance of $H$. The next step is to define a
random process $\{W_t\}$ by
\begin{eqnarray}
W_t = \xi_t-\int_0^t\sigma_sH_s\rd s. \label{eq:5.3}
\end{eqnarray}
It follows then from (\ref{eq:5.1}) that the dynamical equation
for $\{H_t\}$ is given by:
\begin{eqnarray}
\rd H_t= \sigma_t V_t \rd W_t. \label{eq:5.4}
\end{eqnarray}
Equation (\ref{eq:5.4}) can be given a simple heuristic
interpretation if we write it in the form
\begin{eqnarray}
H_{t+{\rm d}t} - {\mathbb E}\left[H_{t+{\rm d}t}|\{\xi_s\}_{0\leq
s\leq t}\right]  = \sigma_t V_t \rd W_t .
\end{eqnarray}
Since at $t$ the variance $V_t$ is known, as is also the
conditional expectation ${\mathbb E}[H_{t+{\rm
d}t}|\{\xi_s\}_{0\leq  s\leq t}]$, we see that $\rd W_t$ embodies
the ``new information'' entering the system between $t$ and $t+\rd
t$. It is for this reason that $\{ W_t\}$ is called an
\emph{innovation process}.

We claim that $\{W_t\}$ is an $\{{\mathcal F}_t\}$-Brownian
motion. Here we recall that $\{{\mathcal F}_t\}$ denotes the
filtration generated by the process $\{\xi_t\}$. Thus,
conditioning with respect to the $\sigma$-algebra ${\mathcal F}_t$
means conditioning with respect to the trajectory
$\{\xi_s\}_{0\leq s\leq t}$. To proceed we need now a more precise
definition of Brownian motion. A process $\{W_t\}_{0\leq t\leq
\infty}$ on a probability space $(\Omega,{\mathcal F},{\mathbb
P})$ with filtration $\{{\mathcal F}_t\}_{0\leq t\leq \infty}$ is
said to be a standard Brownian motion if it satisfies the
following properties: (i) $W_0=0$ almost surely; (ii) $\{W_t\}$ is
$\{{\mathcal F}_t\}$-adapted; and (iii) for all $0\leq s\leq t$
the increment $W_t-W_s$ is normally distributed with mean zero and
variance $t-s$, and is independent of ${\mathcal F}_s$. In the
present context we shall use the so-called L\'evy's
characterisation of Brownian motion, which states that if
$\{W_t\}$ is a martingale, and if $(\rd W_t)^2=\rd t$, then
$\{W_t\}$ is a Brownian motion.

Let us consider the martingale condition. Writing ${\mathbb
E}_t[-]={\mathbb E} [-|{\mathcal F}_t]$ for conditional
expectation with respect to ${\mathcal F}_t$, we shall establish
that ${\mathbb E}_t[W_T] = W_t$ for $t\leq T$. We find
\begin{eqnarray}
{\mathbb E}_t\left[W_T\right]&=& {\mathbb E}_t\left[\xi_T\right] -
{\mathbb E}_t\left[\int_0^T\!\sigma_s H_s \rd s\right] \nonumber \\
&=& H_t\int_0^T\!\sigma_s\rd s +{\mathbb E}_t\left[ B_T\right] -
\int_0^T\!\sigma_s {\mathbb E}_t\left[ H_s \right] {\rd}s,
\label{eq:5.5}
\end{eqnarray}
where we have substituted (\ref{eq:1.3}) and we have interchanged
the order of integration and expectation by use of the Fubini
theorem. Next, we note that
\begin{eqnarray}
\int_0^T\!\sigma_s {\mathbb E}_t\left[H_s\right]{\rd}s &=&
\int_0^t\!\sigma_s {\mathbb E}_t\left[H_s\right]{\rd}s + \int_t^T
\!\sigma_s {\mathbb E}_t\left[H_s\right]{\rd}s \nonumber \\ &=&
\int_0^t\!\sigma_s H_s {\rm d}s + H_t \int_t^T\!\sigma_s{\rd}s.
\label{eq:5.6}
\end{eqnarray}
Here we have used the fact that $H_t={\mathbb E}_t[H]$ satisfies
the martingale condition ${\mathbb E}_t [H_s]=H_t$ for $s\geq t$.
Hence substituting (\ref{eq:5.6}) into (\ref{eq:5.5}) we obtain
\begin{eqnarray}
{\mathbb E}_t\left[W_T\right] = H_t\int_0^t\!\sigma_s\rd s +
{\mathbb E}_t\left[B_T\right] - \int_0^t\!\sigma_s H_s{\rd}s .
\label{eq:5.7}
\end{eqnarray}
Finally, from the tower property of conditional expectation we
have
\begin{eqnarray}
{\mathbb E}_t\left[B_T\right] = {\mathbb E}_t\Big[ {\mathbb E}
\left[B_T| \{B_s\}_{0\leq s\leq t},H\right]\Big]={\mathbb E}_t
\left[B_t\right] . \label{eq:5.8}
\end{eqnarray}
Inserting this relation into (\ref{eq:5.7}) we obtain
\begin{eqnarray}
{\mathbb E}_t\left[W_T\right] &=& {\mathbb E}_t\left[H\int_0^t
\sigma_s \rd s + B_t \right] - \int_0^t\!\sigma_s H_s{\rd}s
\nonumber \\ &=& \xi_t- \int_0^t\!\sigma_s H_s{\rd}s  = W_t,
\label{eq:5.9}
\end{eqnarray}
where we have used the relation ${\mathbb E}_t[\xi_t]=\xi_t$. This
establishes that $\{W_t\}$ is an $\{{\mathcal F}_t\}$-martingale. On
the other hand, because
\begin{eqnarray}
{\rd}W_t=(H-H_t)\sigma_t{\rd}t+{\rd}B_t, \label{eq:5.10}
\end{eqnarray}
it follows that $({\rd}W_t)^2={\rd}t$. Taking this together with
the fact that $\{W_t\}$ is an $\{{\mathcal F}_t\}$-martingale, we
conclude by L\'evy's criterion that $\{W_t\}$ is an $\{{\mathcal
F}_t\}$-Brownian motion.

We are now closer to establishing the relation between
(\ref{eq:1.2}) and (\ref{eq:3.7}). To this end we consider the
conditional probability (\ref{eq:2.26a}) that $H$ takes the value
$E_i$. Taking the stochastic differential of (\ref{eq:2.26a}) and
substituting (\ref{eq:5.3}) into the result, we find, after some
rearrangement, that $\{\pi_{it}\}$ satisfies
\begin{eqnarray}
\rd\pi_{it}=\sigma_t(E_i-H_t)\pi_{it}\rd W_t. \label{eq:5.12}
\end{eqnarray}
With another application of the Ito formula, we thus deduce that
\begin{eqnarray}
\rd\pi_{it}^{1/2} = - \octa\sigma_t^2(E_i-H_t)^2\pi_{it}^{1/2} \rd
t + \half \sigma_t(E_i-H_t)\pi_{it}^{1/2} \rd W_t. \label{eq:5.13}
\end{eqnarray}
Finally, if we let $|\phi_i\rangle$ denote the normalised L\"uders
state~\cite{abbh,luders} associated with the eigenvalue $E_i$, and
define $\{|\psi_t\rangle \}$ according to
\begin{eqnarray}
|\psi_t\rangle = \sum_{i} {\re}^{-{\rm i}E_i t} \pi_{it}^{1/2}
|\phi_i\rangle, \label{eq:5.14}
\end{eqnarray}
then it follows at once from (\ref{eq:5.13}) that
$\{|\psi_t\rangle\}$ satisfies the time-dependent energy-based
stochastic Schr\"odinger equation (\ref{eq:1.1}).

In summary, if we define $\{\xi_t\}$ and $\{\eta_t\}$ in terms of
the independent random data $H$ and $\{B_t\}$ according to
(\ref{eq:1.3}) and (\ref{eq:3.6}), and if we define $\{H_t\}$ and
$\{|\psi_t\rangle\}$ by (\ref{eq:3.7}) and (\ref{eq:5.14}), and
$\{W_t\}$ by (\ref{eq:5.3}), then $\{|\psi_t\rangle\}$ solves
(\ref{eq:1.1}) for the given Hamiltonian ${\hat H}$ and initial
condition $|\psi_0\rangle$.

\section{Direct derivation of independent random data}
\label{sec:65}

The way in which we have solved equation (\ref{eq:1.1}) is by
introducing the concept of a quantum information process
(\ref{eq:1.3}) specified in terms of a random variable $H$ and an
independent Brownian motion $\{B_t\}$. It is possible, however, to
\textit{deduce} the existence of these random data directly from
(\ref{eq:1.1}). In this section we shall illustrate this reverse
construction.

We begin by remarking that the energy-based stochastic
Schr\"odinger equation can be cast into integral form,
incorporating the initial condition $|\psi_0\rangle$, as follows:
\begin{eqnarray}
|\psi_t\rangle = \exp\left( -{\ri}{\hat H}t - \quat \int_0^t
\sigma_s^2 \big({\hat H}-H_s\big)^2{\rd}s + \half \int_0^t
\sigma_s \big({\hat H}-H_s\big) {\rd}W_s \right) |\psi_0\rangle .
\end{eqnarray}
After some simple rearrangement we then deduce that
\begin{eqnarray}
|\psi_t\rangle = \frac{\exp\left( -{\rm i}{\hat H}t + \frac{1}{2}
{\hat H} \int_0^t \sigma_s ({\rm d}W_s+\sigma_s H_s{\rm d}s) -
\frac{1}{4} {\hat H}^2 \int_0^t \sigma_s^2 {\rm d}s
\right)|\psi_0\rangle} {\exp\left( \frac{1}{2} \int_0^t \sigma_s
H_s ({\rm d} W_s+\sigma_s H_s{\rm d}s) - \frac{1}{4} \int_0^t
\sigma_s^2 H_s^2 {\rm d}s \right)} . \label{eq:10.1}
\end{eqnarray}
Given the $\{{\mathcal F}_t\}$-adapted Brownian motion $\{W_t\}$
and the energy expectation process $\{H_t\}=\langle\psi_t|{\hat H}
|\psi_t\rangle$ we now \textit{define} a process $\{\xi_t\}$ by
writing
\begin{eqnarray}
\xi_t = W_t + \int_0^t\sigma_sH_s\rd s. \label{eq:10.2}
\end{eqnarray}
It follows then that $|\psi_t\rangle$ can be written in the form
\begin{eqnarray}
|\psi_t\rangle = \frac{\exp\left( -{\rm i}{\hat H}t + \frac{1}{2}
{\hat H} \int_0^t \sigma_s {\rm d}\xi_t - \frac{1}{4} {\hat H}^2
\int_0^t \sigma_s^2 {\rm d}s \right)|\psi_0\rangle} {\exp\left(
\frac{1}{2} \int_0^t \sigma_s H_s {\rm d}\xi_t - \frac{1}{4}
\int_0^t \sigma_s^2 H_s^2 {\rm d}s \right)} . \label{eq:10.1.5}
\end{eqnarray}

With these ingredients at hand we now claim the following:
\textit{The random variables $H=\lim_{t\to\infty}H_t$ and $B_t =
\xi_t -H \int_0^t \sigma_s \rd s$ are independent. Furthermore,
the process $\{B_t\}$ thus defined is a standard ${\mathbb
P}$-Brownian motion}.

The existence of the random variable $H$ is ensured by the
martingale convergence theorem. The fact that $H$ has the
distribution ${\mathbb P}(H=E_i)=\pi_i$ then follows as a
consequence of known properties of the stochastic equation
(\ref{eq:1.1}). To show that $H$ and $B_t$ are independent (for
any value of $t$) it suffices to show that
\begin{eqnarray}
{\mathbb E}^{\mathbb P}\left[\re^{x B_t+y H}\right] ={\mathbb
E}^{\mathbb P}\left[\re^{x B_t}\right] {\mathbb E}^{\mathbb P}
\left[\re^{y H}\right]
\end{eqnarray}
for any $x,y$. The proof proceeds as follows. First, by use of the
tower property of conditional expectation we have
\begin{eqnarray}
{\mathbb E}^{\mathbb P}\left[\re^{x B_t+y H}\right] &=& {\mathbb
E}^{\mathbb P}\left[\re^{x (\xi_t-H\int_0^t \sigma_s {\rm d}s)+y
H}\right] \nonumber \\ &=& {\mathbb E}^{\mathbb P} \left[\re^{x
\xi_t}\,{\mathbb E}_t^{\mathbb P} \left[ \re^{(y-x\int_0^t
\sigma_s {\rm d}s) H}\right] \right], \label{eq:inner}
\end{eqnarray}
where ${\mathbb E}_t$ denotes conditional expectation with respect
to ${\mathcal F}_t$. Here we have used the fact that $\xi_t$ is
${\mathcal F}_t$-measurable. For the inner expectation we can
write
\begin{eqnarray}
{\mathbb E}_t^{\mathbb P}\left[ \re^{(y-x\int_0^t \sigma_s {\rm
d}s) H}\right] = \sum_i \pi_{it}\,\re^{(y-x\int_0^t \sigma_s {\rm
d}s) E_i},  \label{eq:10.5}
\end{eqnarray}
where $\pi_{it}$ is defined as the conditional probability
\begin{eqnarray}
\pi_{it}={\mathbb P}\left( H=E_i|{\mathcal F}_t\right).
\end{eqnarray}
To obtain an expression for $\{\pi_{it}\}$ we recall \cite{ah,
abbh} that since the projection operator ${\hat\Pi}_i$ commutes
with the Hamiltonian, the bounded process $\{\langle\psi_t|{\hat
\Pi}_i| \psi_t\rangle\}$ is a martingale. It follows that
\begin{eqnarray}
\langle\psi_t|{\hat\Pi}_i| \psi_t\rangle = {\mathbb E}_t^{\mathbb
P}\left[ \langle\psi_\infty|{\hat\Pi}_i| \psi_\infty\rangle
\right] .
\end{eqnarray}
On the other hand, by known properties of the reduction process
(\ref{eq:1.1}) we have $\langle\psi_\infty|{\hat\Pi}_i|
\psi_\infty\rangle = {\mathbf 1}_{ \{H=E_i\}}$, and hence
\begin{eqnarray}
\langle\psi_t|{\hat\Pi}_i| \psi_t\rangle = {\mathbb E}_t^{\mathbb
P} \left[ {\mathbf 1}_{ \{H=E_i\}} \right] .
\end{eqnarray}
We are therefore able to deduce that ${\mathbb P}(H=E_i|{\mathcal
F}_t) = \pi_{it}=\langle\psi_t|{\hat\Pi}_i| \psi_t\rangle$. A
short calculation making use of (\ref{eq:10.1.5}) and properties
of the projection operator then shows that
\begin{eqnarray}
\pi_{it} = \frac{\pi_i \exp\left( E_i \int_0^t \sigma_s {\rm d}
\xi_t - \frac{1}{2} E_i^2\int_0^t \sigma_s^2 {\rm d}s \right)}
{\exp\left( \int_0^t \sigma_s H_s {\rm d}\xi_t - \frac{1}{2}
\int_0^t \sigma_s^2 H_s^2 {\rm d}s \right)} . \label{eq:10.1.55}
\end{eqnarray}

In fact, with a little work one can also show that $\pi_{it}=
|\langle\phi_i|\psi_t\rangle|^2$. In other words, $\pi_{it}$ is
given by the usual formula for the quantum-mechanical transition
probability from the state $|\psi_t\rangle$ to the L\"uders state
$|\phi_i\rangle$. Of course, in standard quantum mechanics it is
an \textit{assumption} that $\pi_{it}$, when defined in this way,
has the interpretation of a transition probability. But in the
stochastic theory, we deduce this property.

Returning to our calculation of the inner conditional expectation
in formula (\ref{eq:inner}) we see as a consequence of
(\ref{eq:10.1.55}) that
\begin{eqnarray}
{\mathbb E}_t^{\mathbb P} \left[ \re^{(y-x\int_0^t \sigma_s {\rm
d}s) H}\right] = \frac{\sum_i \pi_i\, \re^{(y-x\int_0^t \sigma_s
{\rm d}s) E_i} \,\re^{ E_i \int_0^t \sigma_s {\rm d} \xi_t -
\frac{1}{2} E_i^2\int_0^t \sigma_s^2 {\rm d}s }} {\re^{ \int_0^t
\sigma_s H_s {\rm d}\xi_t - \frac{1}{2} \int_0^t \sigma_s^2 H_s^2
{\rm d}s }}. \label{eq:inner2}
\end{eqnarray}
It follows that
\begin{eqnarray}
{\mathbb E}^{\mathbb P} \left[ \re^{xB_t+yH} \right] &=& {\mathbb
E}^{\mathbb P} \left[ \frac{\re^{x\xi_t} \sum_i \pi_i\, \re^{(y-x
\int_0^t \sigma_s {\rm d}s) E_i} \,\re^{ E_i \int_0^t \sigma_s
{\rm d} \xi_t - \frac{1}{2} E_i^2\int_0^t \sigma_s^2 {\rm d}s }}
{\re^{ \int_0^t \sigma_s H_s {\rm d}\xi_t - \frac{1}{2} \int_0^t
\sigma_s^2 H_s^2 {\rm d}s }} \right] \nonumber
\\ &=&
\sum_i \pi_i\,\re^{(y-x\int_0^t \sigma_s {\rm d}s) E_i}\, {\mathbb
E}^{\mathbb P} \left[ \frac{\re^{x\xi_t}\, \re^{ E_i \int_0^t
\sigma_s {\rm d} \xi_t - \frac{1}{2} E_i^2\int_0^t \sigma_s^2 {\rm
d}s }} {\re^{ \int_0^t \sigma_s H_s {\rm d}\xi_t - \frac{1}{2}
\int_0^t \sigma_s^2 H_s^2 {\rm d}s }} \right] . \label{eq:inner3}
\end{eqnarray}
Next we observe that since $\rd\xi_t=\rd W_t+\sigma_tH_t\rd t$,
the expectation appearing above can be written in the form
\begin{eqnarray}
{\mathbb E}^{\mathbb P} \left[ \frac{\re^{x\xi_t}\, \re^{ E_i
\int_0^t \sigma_s {\rm d} \xi_t - \frac{1}{2} E_i^2\int_0^t
\sigma_s^2 {\rm d}s }} {\re^{ \int_0^t \sigma_s H_s {\rm d}\xi_t -
\frac{1}{2} \int_0^t \sigma_s^2 H_s^2 {\rm d}s }} \right]=
{\mathbb E}^{\mathbb P} \left[ \re^{-\int_0^t \sigma_s H_s {\rm d}
W_s - \frac{1}{2}\int_0^t \sigma_s^2 H_s^2 {\rm d}s}\,
\re^{x\xi_t}\, \re^{ E_i \int_0^t \sigma_s {\rm d} \xi_t -
\frac{1}{2} E_i^2\int_0^t \sigma_s^2 {\rm d}s }\right].
\end{eqnarray}
However, the expression
\begin{eqnarray}
\Phi_t = \exp\left( -\int_0^t \sigma_s H_s {\rm d}W_t - \half
\int_0^t \sigma_s^2 H_s^2 {\rm d}s \right)
\end{eqnarray}
is the change-of-measure density over the interval $[0,t]$ needed
to make $\{\xi_s\}_{0\leq s\leq t}$ a standard Brownian motion.
Writing ${\mathbb Q}$ for the resulting new measure, by use of the
Girsanov theorem we obtain
\begin{eqnarray}
{\mathbb E}^{\mathbb P} \left[ \re^{xB_t+yH} \right] = \sum_i
\pi_i\,\re^{(y-x\int_0^t \sigma_s {\rm d}s) E_i}\, {\mathbb
E}^{\mathbb Q} \left[ \re^{x\xi_t}\, \re^{ E_i \int_0^t \sigma_s
{\rm d} \xi_t - \frac{1}{2} E_i^2\int_0^t \sigma_s^2 {\rm d}s }
\right]. \label{eq:inner33}
\end{eqnarray}
By rearranging terms, the expectation on the right above can be
rewritten in the form
\begin{eqnarray}
{\mathbb E}^{\mathbb Q} \left[ \re^{x\xi_t}\, \re^{ E_i \int_0^t
\sigma_s {\rm d} \xi_t - \frac{1}{2} E_i^2\int_0^t \sigma_s^2 {\rm
d}s } \right] = \re^{\frac{1}{2}x^2 t+x E_i \int_0^t \sigma_s {\rm
d}s}\, {\mathbb E}^{\mathbb Q} \left[\re^{ \int_0^t (E_i\sigma_s+
x) {\rm d} \xi_t - \frac{1}{2} \int_0^t (E_i\sigma_s+x)^2 {\rm d}s
} \right] . \label{eq:inner333}
\end{eqnarray}
Since $\{\xi_t\}$ is a ${\mathbb Q}$-Brownian motion, it follows
that the expectation appearing on the right side of
(\ref{eq:inner333}) has the value unity. As a consequence, we
deduce that
\begin{eqnarray}
{\mathbb E}^{\mathbb P} \left[ \re^{xB_t+yH} \right] =
\re^{\frac{1}{2}x^2t}\,\sum_i \pi_i\,\re^{yE_i} .
\end{eqnarray}
We see therefore that $B_t$ and $H$ are independent, as claimed.
We also see, by the form of its characteristic function, that
$B_t$ is Gaussian, with mean 0 and variance $t$. An argument
similar to that presented above shows that if $s\leq t$ then
\begin{eqnarray}
{\mathbb E}^{\mathbb P} \left[ \re^{xB_s+y(B_t-B_s)} \right] =
\re^{\frac{1}{2}x^2s}\,\re^{y^2(t-s)}
\end{eqnarray}
for all $x,y$, and hence ${\rm Cov}(B_s,B_t)=s$ for $s\leq t$.
Since $\{B_t\}$ is a continuous Gaussian process and has the
correct mean and autocovariance properties, we deduce that
$\{B_t\}$ is a standard ${\mathbb P}$-Brownian motion.

We have therefore shown that the solution to (\ref{eq:1.1}) can be
put into the form (\ref{eq:10.1.5}), with $\xi_t=H\int_0^t
\sigma_s \rd s+B_t$, where $H=\lim_{t\to\infty} \langle\psi_t|
{\hat H}|\psi_t\rangle$, and where $\{B_t\}$ is an independent
standard Brownian motion.

Conversely, as we have shown earlier in the paper, if on a fixed
probability space $(\Omega,{\mathcal F}, {\mathbb P})$ we are
given a random variable $H$ with the distribution ${\mathbb
P}(H=E_i)=\pi_i$ together with an independent standard Brownian
motion $\{B_t\}$, then we can proceed as follows: we construct the
process $\{\xi_t\}$ by setting $\xi_t=H\int_0^t\sigma_s\rd s+
B_t$; we define the process $\{H_t\}$ by $H_t={\mathbb E}[H|
{\mathcal F}_t]$ where $\{{\mathcal F}_t\}$ is the filtration
generated by $\{\xi_t\}$; we define the process $\{\pi_{it}\}$ by
setting $\pi_{it}={\mathbb E}[{\mathbf 1}_{\{H=E_i\}}| {\mathcal
F}_t]$; and we define the $\{{\mathcal F}_t\}$-adapted Brownian
motion $\{W_t\}$ by setting $W_t=\xi_t-\int_0^t\sigma_sH_s\rd s$.
Then $\{|\psi_t\rangle\}$, defined by $|\psi_t\rangle=\sum_i
\re^{-{\rm i}E_it} \pi_{it}^{1/2} |\phi_i\rangle$, satisfies
(\ref{eq:1.1}).

Let us now consider briefly the ``linearisation'' technique (see,
e.g., \cite{bassi} and references cited therein) often used for
studying the dynamics of (\ref{eq:1.1}). Starting with
(\ref{eq:1.1}), we proceed as above to deduce (\ref{eq:10.1.5}),
defining the process $\{\xi_t\}$ as in (\ref{eq:10.2}). Next we
observe that if we introduce a process $\{|\Psi_t\rangle\}$ by
defining
\begin{eqnarray}
|\Psi_t\rangle = \exp\left( -{\rm i}{\hat H}t + \half {\hat H}
\int_0^t \sigma_s {\rm d}\xi_t - \quat {\hat H}^2 \int_0^t
\sigma_s^2 {\rm d}s \right)|\psi_0\rangle, \label{eq:Psi}
\end{eqnarray}
then $|\Psi_t\rangle$ satisfies
\begin{eqnarray}
{\rm d}|\Psi_t\rangle = -{\ri}{\hat H} |\Psi_t\rangle {\rm d}t -
\octa \sigma_t^2{\hat H}^2 |\Psi_t\rangle {\rd}t + \half \sigma_t
{\hat H}|\Psi_t\rangle {\rd}\xi_t . \label{eq:10.3}
\end{eqnarray}
We note that $|\Psi_t\rangle$ appears in the numerator of
(\ref{eq:10.1.5}). Thus the relation between $|\Psi_t\rangle$ and
$|\psi_t\rangle$ is given by
\begin{eqnarray}
|\psi_t\rangle = \frac{|\Psi_t\rangle}{\sqrt{\langle\Psi_t|
\Psi_t\rangle}}, \label{eq:Psi2}
\end{eqnarray}
where
\begin{eqnarray}
\langle\Psi_t|\Psi_t\rangle = \exp\left(\int_0^t \sigma_s H_s {\rm
d}\xi_t - \half \int_0^t \sigma_s^2 H_s^2 {\rm d}s \right).
\end{eqnarray}

Now suppose we fix a finite time interval $[0,T]$, and change to a
new measure ${\mathbb Q}$ by use of the density $\Phi_T =
\langle\Psi_T|\Psi_T\rangle^{-1}$. Then over the interval $[0,T]$
the process $\{\xi_t\}$ is a ${\mathbb Q}$-Brownian motion. Thus
in the measure ${\mathbb Q}$ we can ``solve'' (\ref{eq:1.1}) by
expressing $|\psi_t\rangle$ in terms of $|\Psi_t\rangle$, making
use of (\ref{eq:Psi}) and (\ref{eq:Psi2}). Equation
(\ref{eq:10.3}) holds under both ${\mathbb P}$ and ${\mathbb Q}$,
but under ${\mathbb Q}$ it is a linear equation, and hence under
${\mathbb Q}$ we can regard (\ref{eq:Psi}) as the solution of
(\ref{eq:10.3}).

The existence of this underlying ``linearisation'' of
(\ref{eq:1.1}), together with the fact that there is only a single
Hermitian operator appearing in the dynamics, may help to explain
why the problem is exactly solvable. Of course, the ``physics'' is
in the measure ${\mathbb P}$, so any application of the
linearisation technique to solve a physical model typically
involves using the density $\Phi_T^{-1}$ to change from ${\mathbb
Q}$ back to ${\mathbb P}$. In practical terms this means that
realistic simulations of the trajectories of $|\psi_t\rangle$
cannot be efficiently achieved by use of the linearisation
technique.

On the other hand, our ``random data'' method involves a
construction that is carried out entirely in the physical measure.
The auxiliary measures are introduced simply for the purpose of
verifying the results, not for the actual specification of the
solution. In that sense, the random data method can be regarded as
a major improvement over the linearisation method. In particular,
to simulate a set of trajectories for $\{|\psi_t\rangle\}$ we only
need to simulate outcomes for $H$ and $\{B_t\}$ in the measure
${\mathbb P}$.

\section{Verification of collapse property}
\label{sec:55}

In this section we shall verify directly that the solution
(\ref{eq:5.14}) of the stochastic Schr\"odinger equation
(\ref{eq:1.1}) gives rise to the collapse of the wave function. By
substituting (\ref{eq:1.3}) into (\ref{eq:2.26a}), setting
$H=E_k$, and inserting the resulting expression into
(\ref{eq:5.14}), we can express the solution of (\ref{eq:1.1}),
conditional on $H=E_k$ for some fixed value of $k$, in the form
\begin{eqnarray}
|\psi_t\rangle = \frac{\sum_i\sqrt{\pi_i}\exp\left(-{\ri}E_i t +
\half E_iE_k \int_0^t\sigma_s^2\rd s + \half E_i \int_0^t\sigma_s
\rd B_s -\quat E_i^2 \int_0^t\sigma_s^2\rd s \right)|\phi_i\rangle}
{\left[\sum_i \pi_i \exp\left(E_iE_k \int_0^t\sigma_s^2\rd s + E_i
\int_0^t\sigma_s \rd B_s - \half E_i^2 \int_0^t\sigma_s^2\rd s
\right) \right]^{1/2}}. \label{eq:5.15}
\end{eqnarray}
In this situation we imagine that nature has ``secretly'' chosen
the outcome $H=E_k$ (i.e. $H$ takes this value for the given
$\omega\in\Omega$), and we want to show that the wave function
evolves to the appropriate eigenstate. If we multiply the
numerator and denominator of (\ref{eq:5.15}) by $\exp( -\quat
E_k^2 \int_0^t\sigma_s^2\rd s - \half E_k \int_0^t\sigma_s \rd
B_s)$ and write $\omega_{ik}=E_i-E_k$, then (\ref{eq:5.15})
becomes
\begin{eqnarray}
|\psi_t\rangle&=&\frac{\sum_i\sqrt{\pi_i}\exp\left(-{\ri}E_i t -
\quat \omega_{ik}^2 \int_0^t\sigma_s^2\rd s + \half \omega_{ik}
\int_0^t\sigma_s \rd B_s \right)|\phi_i\rangle}{\left[\sum_i \pi_i
\exp\left(-\half \omega_{ik}^2 \int_0^t\sigma_s^2\rd s + \omega_{ik}
\int_0^t\sigma_s \rd B_s \right) \right]^{1/2}} \nonumber \\ &=&
\frac{\sqrt{\pi_k}{\re}^{-{\rm i}E_k t} |\phi_k\rangle + \sum_{i\neq
k}\sqrt{\pi_i}\exp\left( -{\ri}E_i t-\quat \omega_{ik}^2 \int_0^t
\sigma_s^2\rd s + \half \omega_{ik} \int_0^t\sigma_s \rd B_s \right)
|\phi_i\rangle} {\left[\pi_k+\sum_{i\neq k}\pi_i \exp\left(-\half
\omega_{ik}^2 \int_0^t\sigma_s^2\rd s + \omega_{ik} \int_0^t\sigma_s
\rd B_s \right) \right]^{1/2}} . \label{eq:5.16}
\end{eqnarray}
It should be evident then, on account of condition (\ref{eq:1.4}),
that $|\psi_t\rangle\to{\re}^{-{\rm i}E_k t}|\phi_k\rangle$ as
$t\to\infty$. More precisely, defining
\begin{eqnarray}
M_t=\exp\left(\half \omega \int_0^t\sigma_s \rd B_s -\quat
\omega^2 \int_0^t \sigma_s^2\rd s\right),
\end{eqnarray}
we have, for any $\epsilon>0$,
\begin{eqnarray}
{\mathbb P}(M_t>\epsilon) &=& {\mathbb P}\left( \half\omega
\int_0^t\sigma_s \rd B_s -\quat \omega^2 \int_0^t \sigma_s^2\rd
s>\ln\epsilon\right) \nonumber \\
&=& {\mathbb P}\left( \frac{\int_0^t\sigma_s \rd B_s}{
\sqrt{\int_0^t \sigma_s^2\rd s}} > \half \omega \sqrt{\int_0^t
\sigma_s^2\rd s} + \frac{2\ln\epsilon}{\omega\sqrt{\int_0^t
\sigma_s^2\rd s}}\right) \nonumber \\
&=& \left\{ \begin{array}{ll} 1-N\left(\half \omega \sqrt{\int_0^t
\sigma_s^2\rd s} + \frac{2\ln\epsilon}{\omega\sqrt{\int_0^t
\sigma_s^2\rd s}}\right) & (\omega>0) \\
N\left( -\half \omega \sqrt{\int_0^t \sigma_s^2\rd s} -
\frac{2\ln\epsilon}{\omega\sqrt{\int_0^t \sigma_s^2\rd s}}\right) &
(\omega<0) ,
\end{array} \right.
\end{eqnarray}
where $N(x)$ is the standard normal distribution function
\begin{eqnarray}
N(x)=\frac{1}{\sqrt{2\pi}}\int_{-\infty}^x\re^{-\frac{1}{2}y^2}\rd
y.
\end{eqnarray}
Here we have used the fact that $\int_0^t\sigma_s \rd B_s/(
\int_0^t \sigma_s^2\rd s )^{1/2}$ is normally distributed with
mean zero and variance unity. We thus see that (\ref{eq:1.4}) is
satisfied if and only if ${\mathbb P}(M_t>\epsilon)\to0$ as
$t\to\infty$; and hence it follows that, given condition
(\ref{eq:1.4}), the state vector collapses to the designated
eigenstate. The intuition behind this result is that the
``signal'' component of $\{\eta_t\}$ eventually dominates over the
``noise'' component if (\ref{eq:1.4}) is satisfied. This is
because the magnitude of $\int_0^t\sigma_s \rd B_s$ is on average
about $(\int_0^t \sigma_s^2\rd s )^{1/2}$.

We note, incidentally, that if the leading order behaviour of the
integral of $\{\sigma_t^2\}$ is such that $\int_0^t \sigma_s^2 \rd
s\sim t^\alpha$, with $\alpha>0$, then to leading order we have
$\sigma_t\sim t^{\frac{1}{2}(\alpha-1)}$, and hence $\int_0^t
\sigma_s \rd s\sim t^{\frac{1}{2}(\alpha+1)}$. Since the magnitude
of $B_t$ is on average of the order $t^{1/2}$, we see that in this
situation the signal component in (\ref{eq:1.3}) also dominates over
the noise component.

\section{Reduction without complete collapse}
\label{sec:6}

In this section we show that when condition (\ref{eq:1.4}) is not
satisfied, state reduction nevertheless takes place, in the sense
that the energy variance decreases on average. However, unlike the
models for which (\ref{eq:1.4}) is satisfied, in this case the
terminal energy variance in general does not vanish. In other
words, the state approaches an energy eigenstate, but does not get
there. Physically, this situation corresponds to an
\emph{approximate measurement of energy}, in which some
information concerning the energy of the system is revealed, but
no definite outcome is obtained.

To analyse this situation we consider the energy variance process
(\ref{eq:5.2}), which is given, equivalently, by
\begin{eqnarray}
V_t = \langle{\psi}_t|{\hat H}^2|\psi_t\rangle - \langle{\psi}_t
|{\hat H} |\psi_t\rangle^2. \label{eq:6.1}
\end{eqnarray}
Taking the stochastic differential of (\ref{eq:6.1}) and using the
dynamical equation (\ref{eq:1.1}) we find that
\begin{eqnarray}
{\rd}V_t = -\sigma_t^2 V_t^2 {\rd}t + \sigma_t\kappa_t {\rd}W_t,
\label{eq:6.2}
\end{eqnarray}
where $\kappa_t$ is the third central moment of the energy:
\begin{eqnarray}
\kappa_t = \langle{\psi}_t|({\hat H}-H_t)^3 |\psi_t\rangle.
\label{eq:6.3}
\end{eqnarray}

We observe that the drift of $\{V_t\}$ is strictly negative.
Therefore, the energy variance is on average decreasing. However, if
$\{\sigma_t\}$ is a square-integrable function, then $\{V_t\}$ may
converge to some finite nonzero value smaller than the initial value
$V_0$. To investigate this scenario we write (\ref{eq:6.2}) in
integral form:
\begin{eqnarray}
V_t = V_0 - \int_0^t \sigma_s^2 V_s^2 {\rd}s +\int_0^t\sigma_s
\kappa_s {\rd}W_s. \label{eq:6.4}
\end{eqnarray}
Taking the expectation of each side, we obtain
\begin{eqnarray}
{\mathbb E}\left[V_t\right] = V_0 -\int_0^t \sigma_s^2{\mathbb E}
\left[V_s^2\right]\rd s. \label{eq:6.5}
\end{eqnarray}
On account of Jensen's inequality we have ${\mathbb E}[V_t^2]\geq
({\mathbb E}[V_t])^2$, and hence (\ref{eq:6.5}) implies that
\begin{eqnarray}
{\mathbb E}\left[V_t\right] \leq V_0 -\int_0^t \sigma_s^2\left(
{\mathbb E}[V_s]\right)^2 {\rd}s. \label{eq:6.6}
\end{eqnarray}
Bearing in mind the fact that ${\mathbb E}[V_t]\leq {\mathbb E}
[V_s]$ for $t\geq s$, the inequality (\ref{eq:6.6}) implies
\begin{eqnarray}
{\mathbb E}\left[V_t\right] \leq V_0 -\left( {\mathbb E}
[V_t]\right)^2 \int_0^t \sigma_s^2 {\rd}s. \label{eq:6.7}
\end{eqnarray}
As a consequence, we obtain an upper bound on the expected value of
the energy variance:
\begin{eqnarray}
{\mathbb E}[V_t]\leq \frac{1}{2\int_0^t\sigma_s^2\rd s} \left( -1 +
\sqrt{1+4V_0\int_0^t\sigma_s^2\rd s}\,\right) . \label{eq:6.8}
\end{eqnarray}
In the limit $t\to\infty$ the inequality (\ref{eq:6.8}) determines
an upper bound for the asymptotic value of the expected energy
variance. That is, on average the energy variance will be reduced to
a value no greater than the asymptotic value of the right side of
(\ref{eq:6.8}). In particular, if (\ref{eq:1.4}) is satisfied, then
${\mathbb E}[V_t]\to0$ as $t\to\infty$. On the other hand, if
(\ref{eq:1.4}) is not satisfied, then we can obtain a lower bound
for ${\mathbb E}[V_\infty]$. Let $V_{\rm max}$ denote the maximum
possible variance that the energy can have, over all states. Then
from (\ref{eq:6.5}) we get
\begin{eqnarray}
{\mathbb E}[V_\infty] \geq V_0 - V_{\rm max}^2 \int_0^\infty
\sigma_s^2 \rd s .
\end{eqnarray}
Hence providing $\int_0^\infty \sigma_s^2 \rd s < V_0/V_{\rm max}^2$
we are ensured that state reduction will be incomplete.

\section{Finite-time collapse}
\label{sec:7}

Having investigated the case in which the coupling $\{\sigma_t\}$
decays too rapidly to lead to a complete collapse of the wave
function, we turn to the situation where the integral of the
coupling $\{\sigma_t\}$ diverges over a finite time horizon. In
particular, we consider the example
\begin{eqnarray}
\sigma_t = \frac{\sigma T}{T-t}, \label{eq:7.1}
\end{eqnarray}
where $\sigma>0$ and $T>0$ are fixed constants. For the resulting
state vector dynamics we have the following stochastic
differential equation:
\begin{eqnarray}
\rd |\psi_t\rangle = -\ri{\hat H}|\psi_t\rangle\rd t - \octa
\left(\frac{\sigma T}{T-t}\right)^2 ({\hat H}-H_t )^2|\psi_t
\rangle\rd t +\half \frac{\sigma T}{T-t} ({\hat H}-H_t)
|\psi_t\rangle\rd W_t . \label{eq:7.2}
\end{eqnarray}
This model is of interest because the collapse of the wave
function is achieved in finite time.

The stochastic equation (\ref{eq:7.2}) is identical to the
finite-time collapse model introduced in \cite{bh3}, where a
solution to (\ref{eq:7.2}) is obtained by considering an ansatz of
the form
\begin{eqnarray}
\xi^*_t = \sigma t H + \beta_t . \label{eq:7.3}
\end{eqnarray}
The noise term $\beta_t$ appearing here is a Brownian
bridge~\cite{karatzas,yor} that vanishes at $t=0$ and at $t=T$.
The vanishing of the noise at $t=T$ guarantees the collapse of the
wave function as $t$ approaches $T$. In particular, since the
coupling $\sigma$ in (\ref{eq:7.3}) is constant, $\{\xi_t^*\}$ is
a Markov process, as is shown in \cite{bh4}.

In the present framework, it follows immediately from (\ref{eq:1.3})
that the appropriate ansatz for solving (\ref{eq:7.2}) is given by
\begin{eqnarray}
\xi_t = \sigma T H \ln \Big(\frac{T}{T-t}\Big)+B_t, \label{eq:7.4}
\end{eqnarray}
where $\{B_t\}_{0\leq t\leq T}$ is a standard Brownian motion.
Remarkably, the two prescriptions (\ref{eq:7.3}) and
(\ref{eq:7.4}) give rise to the same solution to the stochastic
equation (\ref{eq:7.2}).

To see this we note that it follows from (\ref{eq:3.7}) that the
energy process associated with (\ref{eq:7.2}) is given by
\begin{eqnarray}
H_t= \frac{\sum_i \pi_i E_i \exp\left(\sigma T E_i \int_0^t
\frac{1}{T-s}\,\rd\xi_s-\half\sigma^2E_i^2 \frac{tT}{T-t}\right)}
{\sum_i\pi_i\exp\left(\sigma TE_i\int_0^t\frac{1}{T-s}\,\rd\xi_s
-\half \sigma^2 E_i^2 \frac{tT}{T-t}\right)} . \label{eq:7.5}
\end{eqnarray}
In Section~\ref{sec:3} we observed that under the probability
measure ${\mathbb Q}$ the process $\{\xi_t\}$ is a Brownian
motion. Therefore, in view of the expression appearing in the
exponent of (\ref{eq:7.5}), we define a process $\{\xi^*_t\}$
according to the following scheme:
\begin{eqnarray}
\xi^*_t = (T-t) \int_0^t \frac{1}{T-s}\,\rd\xi_s . \label{eq:7.6}
\end{eqnarray}
Substituting (\ref{eq:7.4}) into the right side of (\ref{eq:7.6}) we
obtain
\begin{eqnarray}
(T-t) \int_0^t \frac{1}{T-s}\,\rd\xi_s = \sigma TH(T-t)\int_0^t
\frac{1}{(T-s)^2}\,\rd s + (T-t)\int_0^t \frac{1}{T-s}\,\rd B_s.
\label{eq:7.7}
\end{eqnarray}
After a short calculation, we deduce that
\begin{eqnarray}
(T-t) \int_0^t \frac{1}{T-s}\,\rd\xi_s = \sigma t H + \beta_t,
\label{eq:7.75}
\end{eqnarray}
where $\{\beta_t\}$ is defined by
\begin{eqnarray}
\beta_t= (T-t)\int_0^t \frac{1}{T-s}\,\rd B_s . \label{eq:7.8}
\end{eqnarray}
However, we recognise in (\ref{eq:7.8}) a standard integral
representation of a Brownian bridge~\cite{karatzas,yor}. It
follows that $\{\xi_t^*\}$, as defined by (\ref{eq:7.6}), can be
put into the form (\ref{eq:7.3}).

On the other hand, we also see that (\ref{eq:7.6}) is an integral
representation of a Brownian bridge under ${\mathbb Q}$, since in
this measure $\{\xi_t\}$ is a Brownian motion. Therefore, under
${\mathbb Q}$, the energy process (\ref{eq:7.5}) can be expressed
in terms of a Brownian bridge $\{\xi^*_t\}$ in the form
\begin{eqnarray}
H_t = \frac{\sum_i \pi_i E_i \exp\left(\frac{T}{T-t}\left( \sigma
E_i \xi^*_t- \frac{1}{2}\sigma^2 E_i^2 t\right) \right)}{\sum_i
\pi_i\exp\left(\frac{T}{T-t}\left(\sigma
TE_i\xi^*_t-\frac{1}{2}\sigma^2 E_i^2 t\right) \right)} .
\label{eq:7.9}
\end{eqnarray}
This result agrees with the result obtained in \cite{bh3} for the
finite-time collapse model.

We note, incidentally, that the innovation process associated with
$\{\xi_t\}$ in this example, given by
\begin{eqnarray}
W_t = \xi_t - \int_0^t \sigma_s H_s \rd s, \label{eq:7.10}
\end{eqnarray}
where $\sigma_t=\sigma T/(T-t)$, and the innovation process
associated with $\{\xi^*_t\}$, given by
\begin{eqnarray}
W_t = \xi^*_t + \int_0^t \frac{1}{T-s}\Big(\xi^*_s -\sigma T H_s
\Big) \rd s, \label{eq:7.11}
\end{eqnarray}
as obtained in \cite{bh3}, are identical if $\{\beta_t\}$ is
defined as in (\ref{eq:7.8}). This follows on account of the
relation
\begin{eqnarray}
\xi_t=\xi^*_t+\int_0^t\frac{1}{T-s}\,\xi^*_s\rd s, \label{eq:7.12}
\end{eqnarray}
which can be verified by writing the right side of (\ref{eq:7.12})
in differential form:
\begin{eqnarray}
\rd\xi^*_t+\frac{1}{T-t}\,\xi^*_t\rd t&=& \sigma H \rd t +
\rd\beta_t + \frac{1}{T-t}\left(\sigma tH+\beta_t\right)\rd t
\nonumber \\ &=& \sigma H \rd t -\frac{1}{T-t}\,\beta_t\rd t + \rd
B_t + \sigma H \frac{t}{T-t}\,\rd t+\frac{1}{T-t}\,\beta_t\rd t
\nonumber \\ &=& H\frac{\sigma T}{T-t}\,\rd t+\rd B_t = \rd\xi_t.
\label{eq:7.13}
\end{eqnarray}
Therefore, the solution obtained here in terms of $\{\xi_t\}$ is
equivalent to the solution obtained in \cite{bh3} using
$\{\xi^*_t\}$. The results above show that models exhibiting
state-vector reduction over a finite time horizon are both
feasible and tractable, and that such models can be usefully
formulated by use of a time-dependent coupling.

\vskip 15pt \noindent

\begin{acknowledgments}
DCB acknowledges support from The Royal Society; ICC acknowledges
support from UK Particle Physics and Astronomy Research Council;
and JDCD and LPH acknowledge support from the UK Engineering and
Physical Science Research Council (EPSRC). The authors thank
S.L.~Adler, C.M.~Bender, I.R.C.~Buckley, and M.H.A.~Davis for
useful discussions and correspondence, and an anonymous referee
for helpful comments.
\end{acknowledgments}

\appendix

\section{Path integral approach}
\label{app:1}

In this appendix we present a path integral method for calculating
the conditional probability process $\{\pi_{it}\}$ and the energy
expectation process $\{H_t\}$. First we note that (\ref{eq:1.5})
can be written in the form:
\begin{eqnarray}
{\mathbb E}\left[H|\{\xi_s\}_{0\leq s\leq t}\right] = \sum_i E_i
\pi_{it} , \label{eq:1.6}
\end{eqnarray}
where $\pi_{it}={\mathbb P}\left(H=E_i|\{\xi_s\}_{0\leq s\leq
t}\right)$. By use of the Bayes theorem, the conditional
probability is given by
\begin{eqnarray}
{\mathbb P}\left(H=E_i|\{\xi_s\}_{0\leq s\leq t}\right) =
\frac{\pi_i \rho(\{\xi_s\}_{0\leq s\leq t}|H=E_i)} {\sum_i \pi_i
\rho(\{\xi_s\}_{0\leq s\leq t}|H=E_i)}. \label{eq:1.7}
\end{eqnarray}
Here the expression
\begin{eqnarray}
&& \hspace{-0.6cm}\rho(\{\xi_s\}_{0\leq s\leq t}|H=E_i) \nonumber
\\ && = \sqrt{\frac{1}{\det(2\pi\Sigma)}} \exp\left(-\half
\int_0^t\!\!\int_0^t \Sigma^{-1}(u,v) \left(\xi_u-E_i {\textstyle
\int_0^u \sigma_s \rd s}\right)\left(\xi_v-E_i {\textstyle\int_0^v
\sigma_{s} \rd s} \right) \rd u \rd v \right) \label{eq:1.8}
\end{eqnarray}
is the density function of the trajectory $\{\xi_s\}_{0\leq s\leq
t}$ conditional on $H$ taking the value $E_i$. We write
$\Sigma(u,v)$ for the covariance of the random variables
$E_i\int_0^u \sigma_s\rd s+B_u$ and $E_i\int_0^v \sigma_{s} \rd s
+B_v$. The form of the density function (\ref{eq:1.8}) follows
from that fact that, conditional on $H=E_i$, the random variables
$\{\xi_s\}_{0\leq s\leq t}$ are jointly normally distributed. A
straightforward calculation making use of well-known properties of
Brownian motion shows that
\begin{eqnarray}
\Sigma(u,v) = \min(u,v). \label{eq:1.9}
\end{eqnarray}
To compute the inverse $\Sigma^{-1}(u,v)$ of the covariance we
substitute (\ref{eq:1.9}) into the relation
\begin{eqnarray}
\int_0^t \Sigma^{-1}(u,s)\Sigma(s,v) \rd s = \delta(u-v),
\label{eq:1.10}
\end{eqnarray}
and differentiate the resulting expression twice in the variable $v$
to obtain
\begin{eqnarray}
\Sigma^{-1}(u,v)=-\delta^{\prime\prime}(u-v), \label{eq:1.11}
\end{eqnarray}
where $\delta^{\prime\prime}(t)$ denotes the second derivative of
the Dirac delta function.

We note that the conditional expectation of any functional of the
trajectory $\{\xi_s\}_{0\leq s\leq t}$ can be determined by use of
the density function (\ref{eq:1.8}); the resulting expression
corresponds to an infinite-dimensional Feynman integral. This is
related to the fact that $\Sigma(u,v)$ gives the Feynman-Green
function for a free particle~\cite{dewitt,gelfand,xia}.

To determine the form of the density function (\ref{eq:1.8}), one
might consider substituting (\ref{eq:1.11}) into (\ref{eq:1.8}) and
then applying integration by parts, since $-\delta^{\prime\prime}
(t)$ is a second-order differential operator (with the property that
it has a positive ``spike'' at $t=0$ and a pair of negative
``spikes'' at $t=0^+$ and $t=0^-$). However, the integral in the
exponent of (\ref{eq:1.8}) is ill-defined as it stands. Indeed, it
is a straightforward exercise to verify that, depending on the order
in which integration by parts is applied, one obtains different
answers. To circumvent this difficulty we shall discretise the
process $\{\xi_t\}$ first, derive the corresponding expression for
the conditional density function (\ref{eq:1.7}), and then take the
continuum limit.

Our strategy for determining the conditional probability law for the
trajectory $\{\xi_s\}_{0\leq s\leq t}$ is thus as follows. Fixing
$t$, we partition the range $[0,t]$ into $n$ equally-spaced
intervals. In particular, we set $s_k=k\, \Delta$
$(k=0,1,\ldots,n)$, where $\Delta=t/n$. We then calculate the
covariance of the random variables $E_i \int_0^{s_k} \sigma_u\rd
u+B_{s_k}$ and $E_i\int_0^{s_l} \sigma_{u}\rd u+B_{s_l}$, and obtain
following expression for the covariance matrix (cf.
\cite{carmichael}):
\begin{eqnarray}
\Sigma(s_k,s_l) = \left( \begin{array}{ccccccc} s_1 & s_1 & s_1 &
s_1 & \cdots & s_1 & s_1 \\
s_1 & s_2 & s_2 & s_2 & \cdots & s_2 & s_2 \\
s_1 & s_2 & s_3 & s_3 & \cdots & s_3 & s_3 \\
s_1 & s_2 & s_3 & s_4 & \cdots & s_4 & s_4 \\
\vdots & \vdots & \vdots & \vdots & \ddots & \vdots & \vdots \\
s_1 & s_2 & s_3 & s_4 & \cdots & s_{n-1} & s_{n-1} \\
s_1 & s_2 & s_3 & s_4 & \cdots & s_{n-1} & s_n \\
\end{array} \right).
\end{eqnarray}
The validity of this result should be evident from the continuous
case (\ref{eq:1.9}). The inverse of the covariance matrix is the
following second-order difference operator:
\begin{eqnarray}
\left( \begin{array}{cccccc} \frac{1}{s_1-s_0}\!+\!
\frac{1}{s_2-s_1} & -\frac{1}{s_2-s_1} & 0 & 0 & \cdots & 0
\\ -\frac{1}{s_2-s_1} & \frac{1}{s_2-s_1}\!+\!\frac{1}{s_3-s_2} & -
\frac{1}{s_3-s_2} & 0 & \cdots & 0 \\ 0 & -\frac{1}{s_3-s_2} &
\frac{1}{s_3-s_2}\!+\!\frac{1}{s_4-s_3} & -\frac{1}{s_4-s_3} & 0 & 0
\\  \vdots & \ddots & \ddots & \ddots & \ddots & \vdots \\ 0 &
\cdots & \cdots & -\frac{1}{s_{n-1}-s_{n-2}} &
\frac{1}{s_{n-1}-s_{n-2}}\!+\!\frac{1}{s_{n}-s_{n-1}} &
-\frac{1}{s_{n}-s_{n-1}} \\ 0 & \cdots & \cdots & 0 &
-\frac{1}{s_{n}-s_{n-1}} & \frac{1}{s_{n}-s_{n-1}}
\end{array} \right)\!. \label{eq:2.14}
\end{eqnarray}
Note that the last term on the diagonal is anomalous, and is
different from the remaining terms on the diagonal. Because the
partition of the interval $[0,t]$ is equally spaced, we have
$s_k-s_{k-1}=\Delta$ for all $k=1,2,\ldots,n$. Therefore
(\ref{eq:2.14}) simplifies to
\begin{eqnarray}
\Sigma_{kl}^{-1} = \frac{1}{\Delta}\left( \begin{array}{ccccccc}
2 & -1 & 0 & \cdots & \cdots & \cdots & 0 \\
-1 & 2 & -1 & 0 & \cdots & \cdots & 0 \\
0 & -1 & 2 & -1 & 0 & \cdots & 0 \\
\vdots & \ddots & \ddots & \ddots & \ddots & \ddots & \vdots \\
0 & \cdots & 0 & -1 & 2 & -1 & 0  \\
0 & \cdots & \cdots & 0 & -1 & 2 & -1  \\
0 & \cdots & \cdots & \cdots & 0 & -1 & 1  \\
\end{array} \right). \label{eq:2.15}
\end{eqnarray}
For simplicity of notation let us write
\begin{eqnarray}
\alpha_i(s_k) = \xi_{s_k}-E_i\int_0^{s_k}\sigma_u\rd u.
\label{eq:2.16}
\end{eqnarray}
Then the discretised form of the conditional density function
(\ref{eq:1.8}) takes the form
\begin{eqnarray}
\rho\left(\{\xi_{s_1},\xi_{s_2},\ldots,\xi_{s_n}\}|H_T=E_i\right)
&=& \left(\frac{\Delta}{2\pi}\right)^{\frac{1}{2}n} \exp\left(-
\half \sum_{k=1}^n \sum_{l=1}^n \Sigma^{-1}_{kl} \alpha_i(s_k)
\alpha_i(s_l) \right) \nonumber \\ && \hspace{-3.8cm}= \left(
\frac{\Delta}{2\pi}\right)^{\frac{1}{2}n} \exp\left(
\frac{1}{\Delta} \sum_{k=1}^{n-1} \alpha_i(s_k) [\alpha_i
(s_{k+1})-\alpha_i(s_k)] -\frac{1}{2\Delta}\alpha_i^2(s_n) \right).
\label{eq:2.17}
\end{eqnarray}
Here we have used expression (\ref{eq:2.15}) for the inverse
covariance matrix. Let us examine the terms in the exponent.
Substituting definition (\ref{eq:2.16}) we find that
\begin{eqnarray}
\alpha_i(s_k) [\alpha_i (s_{k+1})-\alpha_i(s_k)] &=& \xi_{s_k}(
\xi_{s_{k+1}}-\xi_{s_k})+E_i^2 \int_0^{s_k}\sigma_u \rd u
\int_{s_k}^{s_{k+1}} \sigma_v\rd v \nonumber \\ && \hspace{-3.8cm} +
E_i \left( ( \xi_{s_{k+1}}-\xi_{s_k})\int_{s_k}^{s_{k+1}}
\sigma_u\rd u - \xi_{s_{k+1}} \int_0^{s_{k+1}}\sigma_u \rd u
+\xi_{s_k} \int_0^{s_k}\sigma_u \rd u \right). \label{eq:2.18}
\end{eqnarray}

Turning to the conditional probability (\ref{eq:1.7}) that we aim to
determine, we observe by inspection of the right side of
(\ref{eq:1.7}) that all terms in the exponent that are independent
of the eigenvalue $E_i$, such as the term $\xi_{s_k}( \xi_{s_{k+1}}
-\xi_{s_k})$ in the right side of (\ref{eq:2.18}), cancel. This is
because such terms appear in both the denominator and the numerator.
Hence in what follows we omit such terms. Equality modulo omitted
terms will be denoted by the $\sim$ symbol. Bearing this in mind, we
see that the sum over $k$ of the right side of (\ref{eq:2.18}) gives
\begin{eqnarray}
\sum_{k=1}^{n-1}\alpha_i(s_k) [\alpha_i (s_{k+1})-\alpha_i(s_k)]
&\sim& E_i^2 \sum_{k=1}^{n-1} \int_0^{s_k}\sigma_u \rd u
\int_{s_k}^{s_{k+1}} \sigma_v\rd v \nonumber \\ && \hspace{-3.8cm} +
E_i \left( \sum_{k=1}^{n-1}\sigma_{s_k}(\xi_{s_{k+1}}-\xi_{s_k})
\Delta - \xi_{s_{n}} \int_0^{s_{n}}\sigma_u \rd u +\xi_{s_1}
\sigma_{s_0} \Delta \right). \label{eq:2.19}
\end{eqnarray}
In deducing this result we have used the fact that for sufficiently
large $n$, and hence sufficiently small $\Delta$, the following
relation holds to a high degree of accuracy:
\begin{eqnarray}
\int_{s_k}^{s_{k+1}} \sigma_u\rd u = \sigma_{s_k}\Delta.
\end{eqnarray}
Similarly, omitting the term that contains no $E_i$ we have
\begin{eqnarray}
\alpha_i^2(s_n) \sim -2E_i \xi_{s_n} \int_0^{s_{n}}\sigma_u \rd u +
E_i^2 \left(\int_0^{s_n}\sigma_u\rd u\right)^2. \label{eq:2.21}
\end{eqnarray}
Substituting (\ref{eq:2.19}) and (\ref{eq:2.21}) into
(\ref{eq:2.17}), we deduce that
\begin{eqnarray}
\rho\left(\{\xi_{s_1},\xi_{s_2},\ldots,\xi_{s_n}\}|H_T=E_i\right)
\sim \exp\left(E_i\sum_{k=0}^{n-1}\sigma_{s_k}(\xi_{s_{k+1}} -
\xi_{s_k})-\half E_i^2 \sum_{k=0}^{n-1}\sigma_{s_k}^2\Delta \right).
\label{eq:2.22}
\end{eqnarray}
Here we have made use of the fact that to a high degree of accuracy
one has:
\begin{eqnarray}
\left(\int_0^{s_n}\sigma_u\rd u\right)^2 = \sum_{k=0}^{n-1}
\sigma_{s_k}^2(\Delta t)^2+2\sum_{k\neq l}^{n-1} \sigma_{s_k}
\sigma_{s_l} \Delta^2.
\end{eqnarray}
We have also used the fact that
\begin{eqnarray}
\sum_{k=1}^{n-1}\sum_{l=0}^{k-1} \sigma_{s_k}\sigma_{s_l} =
\sum_{k\neq l}^{n-1}\sigma_{s_k}\sigma_{s_l}.
\end{eqnarray}
Note that the first sum in the exponent of (\ref{eq:2.22}) begins
from $k=0$, and not $k=1$ as in the right side of (\ref{eq:2.19}).
This is because the last term in the right side of (\ref{eq:2.19})
can be written as $\sigma_{s_0} (\xi_{s_1}- \xi_{s_0})\Delta$ (since
$\xi_{s_0}=0$) and thus be absorbed in the sum.

We are now in the position to take the limit $n\to\infty$. In
particular, the first term in the exponent of (\ref{eq:2.22})
converges in this limit to an Ito integral of $\{\sigma_t\}$ with
respect to the process $\{\xi_t\}$, since the discrete approximation
is always taken to be the value of the integrand at $s_k$ in each
interval $[s_k,s_{k+1}]$. The second sum, on the other hand,
converges to the Riemann integral of the function $\{\sigma_t^2\}$.
Writing $\{p_{it}\}$ for the unnormalised density function given by
the right side of (\ref{eq:2.22}), we thus deduce, in the limit
$n\to\infty$, that
\begin{eqnarray}
p_{it} = \exp\left(E_i\int_{0}^{t}\sigma_{s}\rd \xi_{s} -\half E_i^2
\int_{0}^{t}\sigma_{s}^2\rd s\right). \label{eq:2.25}
\end{eqnarray}
As a consequence, for the conditional probability process
$\{\pi_{it}\}$ we obtain (\ref{eq:2.26a}), as desired.

\section{Decomposition of path into increments}
\label{app:2}

In this appendix we introduce another method for obtaining
$\{\pi_{it}\}$, based on the decomposition of the path of the
quantum information process into its increments. The argument goes
as follows. Recalling the dynamical equation (\ref{eq:1.55})
satisfied by the information process $\{\xi_t\}$, we note that
$\{\sigma_t\}$ moderates the strength of the signal showing the
true value of $H$. An equivalent set of information can be
obtained by moderating the noise level by use of the reciprocal
function $\{1/\sigma_t\}$. Thus we consider a process
$\{\zeta_t\}$ defined by
\begin{eqnarray}
\zeta_t = H t + \int_0^t \frac{1}{\sigma_s}\,\rd B_s.
\label{eq:4.1x}
\end{eqnarray}
The relation between $\{\xi_t\}$ and $\{\zeta_t\}$ is given by
\begin{eqnarray}
\rd \zeta_t = \frac{1}{\sigma_t}\,\rd \xi_t. \label{eq:4.2x}
\end{eqnarray}
This alternative representation is motivated by
Wonham~\cite{wonham}, where the filtering equation for a signal
associated with a fixed random variable is investigated.

The fact that the information implicit in $\{\xi_s\}_{0\leq s\leq
t}$ and $\{\zeta_s\}_{0\leq s\leq t}$ is equivalent will be shown.
We begin with the discretisation of the period $[0,t]$ into $n$
equally-spaced intervals of size $\Delta=t/n$, and set
$s_k=k\Delta$ for $k=0,1,\ldots,n$. We define a process for the
increments of $\{\zeta_t\}$ by setting $y_{s_k}=\zeta_{s_{k+1}}-
\zeta_{s_k}$. It follows from (\ref{eq:4.1x}) that
\begin{eqnarray}
y_{s_k}=H\Delta + \int_{s_k}^{s_{k+1}} \sigma_u^{-1}\rd B_u.
\label{eq:4.3x}
\end{eqnarray}
For each value of $n$, the random variables $y_{s_k}-H \Delta$
$(k=0,1,\ldots,n-1)$ are independent and normally distributed, with
mean zero and variance
\begin{eqnarray}
v_{s_k} = \int_{s_k}^{s_{k+1}} \sigma_u^{-2}\rd u . \label{eq:4.4x}
\end{eqnarray}
If we assume that $\Delta$ is small, then to a high degree of
accuracy we have $v_{s_k}=\sigma_{s_k}^{-2}\Delta$. Next we note
that
\begin{eqnarray}
{\mathbb P}\left( H=E_i| \zeta_{s_0},\zeta_{s_1},\cdots,\zeta_{s_n}
\right) = {\mathbb P}\left( H=E_i| y_{s_0},y_{s_1},\cdots,
y_{s_{n-1}} \right). \label{eq:4.5x}
\end{eqnarray}
This relation follows from the fact that conditioning with respect
to $\{\zeta_{s_k}\}_{k=0,1,\ldots,n}$ is equivalent to
conditioning with respect to the corresponding increments. Letting
$\{\pi_{it}^{(n)}\}$ denote the conditional probability defined by
(\ref{eq:4.5x}), we conclude that
\begin{eqnarray}
\pi_{it}^{(n)} &=& \frac{\pi_i\exp\left(-\half\sum
\limits_{k=0}^{n-1} v_{s_k}^{-1}\left(y_{s_k}-E_i\Delta \right)^2
\right)} {\sum_i \pi_i\exp\left(-\half\sum \limits_{k=0}^{n-1}
v_{s_k}^{-1}\left(y_{s_k}-E_i\Delta \right)^2\right)} \nonumber \\
&=& \frac{\pi_i\exp\left( E_i \Delta \sum\limits_{k=0}^{n-1}
v_{s_k}^{-1} y_{s_k} - \half E_i^2
\sum\limits_{k=0}^{n-1}v_{s_k}^{-1} \Delta^2 \right)} {\sum_i
\pi_i\exp\left( E_i \Delta \sum\limits_{k=0}^{n-1}
v_{s_k}^{-1}y_{s_k} - \half E_i^2 \sum\limits_{k=0}^{n-1}
v_{s_k}^{-1} \Delta^2 \right)}. \label{eq:4.6x}
\end{eqnarray}
Substituting the right side of (\ref{eq:4.4x}) into (\ref{eq:4.6x}),
and taking the limit as $n$ gets large, we have
\begin{eqnarray}
\pi_{it} =\lim_{n\to\infty}\pi_{it}^{(n)} =
\frac{\pi_i\exp\left(E_i\int_{0}^{t}\sigma_{s}^2\rd \zeta_{s} -\half
E_i^2 \int_{0}^{t}\sigma_{s}^2\rd s\right)} {\sum_i \pi_i
\exp\left(E_i\int_{0}^{t}\sigma_{s}^2\rd \zeta_{s} -\half E_i^2
\int_{0}^{t}\sigma_{s}^2\rd s\right)}. \label{eq:4.7x}
\end{eqnarray}
In taking the limit to obtain a stochastic integral we have
followed a line of argument similar to that of
Appendix~\ref{app:1}. This verifies that the information content
of the trajectories $\{\xi_s\}_{0\leq s\leq t}$ and
$\{\zeta_s\}_{0\leq s\leq t}$ is the same.


\vspace{0.5cm}

\textbf{References}

\begin{enumerate}

\bibitem{adler} S.~L.~Adler, ``Environmental influence on the
measurement process in stochastic reduction models'' {\em J.
Phys.} A\textbf{35} 841 (2002).

\bibitem{adler2} S.~L.~Adler, ``Weisskopf-Wigner decay theory
for the energy-driven stochastic Schr\"odinger equation'' {\em
Phys. Rev.} D\textbf{67} 25007 (2003).

\bibitem{adler3} S.~L.~Adler, {\em Quantum Theory as an Emergent
Phenomenon} (Cambridge: Cambridge University Press, 2004).

\bibitem{abbh} S.~L.~Adler, D.~C.~Brody, T.~A.~Brun, and
L.~P.~Hughston, ``Martingale models for quantum state reduction''
{\em J. Phys.} A\textbf{34} 8795 (2001).

\bibitem{ab} S.~L.~Adler and T.~A.~Brun,
``Generalized stochastic Schr\"odinger equations for state vector
collapse '' {\em J. Phys.} A\textbf{34} 4797 (2001).

\bibitem{ah} S.~L.~Adler and L.~P.~Horwitz,  ``Structure and
properties of Hughston's stochastic extension of the Schr\"odinger
equation'' {\em J. Math. Phys.} \textbf{41} 2485 (2000).

\bibitem{barchielli} A.~Barchielli and V.~P.~Belavkin
``Measurements continuous in time and \textit{a posteriori} states
in quantum mechanics'' {\em J. Phys.} A\textbf{24} 1495 (1991).

\bibitem{bassi} A.~Bassi and G.~C.~Ghirardi ``Dynamical
reduction models'' {\em Phys. Rep.} \textbf{379} 257 (2003).

\bibitem{bh1} D.~C.~Brody and L.~P.~Hughston, ``Stochastic
reduction in nonlinear quantum mechanics'' {\em Proc. R. Soc.
Lond.} A\textbf{458} 1117 (2002).

\bibitem{bh2} D.~C.~Brody and L.~P.~Hughston, ``Efficient
simulation of quantum state reduction'' {\em J.~Math.~Phys.}
\textbf{43}, 5254 (2002).

\bibitem{bh3} D.~C.~Brody and L.~P.~Hughston, ``Finite-time
stochastic reduction models'' {\em J.~Math.~Phys.} \textbf{46},
082101 (2005).

\bibitem{bh4} D.~C.~Brody and L.~P.~Hughston, ``Quantum noise and
stochastic reduction'' {\em J. Phys.} A\textbf{39} 833 (2006).

\bibitem{bhs} D.~C.~Brody, L.~P.~Hughston, and J.~Syroka
``Relaxation of quantum states under energy perturbations'' {\em
Proc. R. Soc. Lond.} A\textbf{459} 2297 (2003).

\bibitem{bj} R.~S.~Bucy and P.~D.~Joseph, {\em Filtering for
stochastic processes with applications to guidance} (New York:
Interscience Publishers, 1968).

\bibitem{carmichael2} H.~J.~Carmichael, ed., Special issue on
stochastic quantum optics, {\em Quantum and Semiclassical Optics}
\textbf{8}, 49--314 (1996).

\bibitem{carmichael} H.~J.~Carmichael, {\em
Statistical Methods in Quantum Optics 1: Master Equations and
Fokker-Planck Equations} (Berlin: Springer Verlag, 1999).

\bibitem{davis} M.~H.~A.~Davis and S.~I.~Marcus, ``An introduction
to nonlinear filtering'' in {\em Stochastic systems: The
mathematics of filtering and identification and application},
M.~Hazewinkel and J.~C.~Willems, eds. (Dordrecht: D.~Reidel,
1981).

\bibitem{dewitt} C.~DeWitt-Morette, ``Feynman path integrals''
{\em Commun. Math. Phys.} \textbf{37}, 63 (1974).

\bibitem{diosi} L.~Diosi, ``Continuous quantum measurement and
Ito formalism''{\em Phys. Lett.} A\textbf{129} 419 (1988).

\bibitem{gelfand} I.~M.~Gel'fand and A.~M.~Yagrom, ``Integration
in functional spaces and its applications in quantum physics''
{\em J. Math. Phys.} \textbf{1}, 48 (1960).

\bibitem{gpr} G.~C.~Ghirardi, P.~Pearle, and A.~Rimini, ``Markov
processes in Hilbert space and continuous spontaneous localisation
of systems of identical particles'' {\em Phys. Rev.} A\textbf{42},
78 (1990).

\bibitem{gisin} N.~Gisin, ``Stochastic quantum dynamics and
relativity'' {\em Helv. Phys. Acta} \textbf{62}, 363 (1989).

\bibitem{gisin3} N.~Gisin and I.~C.~Percival, ``The quantum-state
diffusion model applied to open systems'' {\em J. Phys.}
A\textbf{25}, 5677 (1992).

\bibitem{hughston}  L.~P.~Hughston, ``Geometry of stochastic
state vector reduction'' {\em Proc. R. Soc. London} A\textbf{452},
953 (1996).

\bibitem{isham} C.~J.~Isham, {\em Lectures on Quantum Theory}
(London: Imperial College Pres, 1995).

\bibitem{kailath} T.~Kailath, ``The structure of Radon-Nikodym
derivatives with respect to Wiener and related measure'' {\em
Ann. Math. Statist.} \textbf{42}, 1054 (1971).

\bibitem{ks} G.~Kallianpur and C.~Striebel, ``Estimation of
stochastic systems: Arbitrary system process with additive white
noise observation errors'' {\em Ann. Math. Statist.} \textbf{39},
785 (1968).

\bibitem{karatzas} I.~Karatzas and S.~E.~Shreve, {\em Brownian
Motion and Stochastic Calculus} (Berlin: Springer, 1997).

\bibitem{ls} R.~S.~Liptser and A.~N.~Shiryaev, {\em Statistics
of Random Processes} Vols. I and II, 2nd ed. (Berlin: Springer
Verlag, 2000).

\bibitem{luders} G.~L\"{u}ders ``\"Uber die
Zustands\"anderung durch den Messprozess'' {\em Ann. Physik}
\textbf{8}, 322 (1951).

\bibitem{pearle} P.~Pearle, `Collapse Models'' in {\em Open
Systems and Measurement in Relativistic Quantum Theory} ed.
H.~P.~Breuer and F.~Petruccione (Heidelberg: Springer, 1999).

\bibitem{pearle1} P.~Pearle, ``Problems and aspects of
energy-driven wave-function collapse models'' {\em Phys. Rev.}
A\textbf{69}, 042106 (2004).

\bibitem{percival} I.~C.~Percival, ``Primary state diffusion''
{\em Proc. R. Soc. Lond.} A\textbf{447}, 189 (1994).

\bibitem{percival2} I.~C.~Percival, {\em Quantum State
Diffusion} (Cambridge: Cambridge University Press, 1998).

\bibitem{wonham} W.~M.~Wonham, ``Some applications of stochastic
differential equations to optimal nonlinear filtering'' {\em J.
SIAM} A\textbf{2}, 347 (1965).

\bibitem{yor} M.~Yor, {\em Some Aspects of Brownian Motion}
Part I~\&~II (Basel: Birkh\"auser Verlag 1992~\&~1996).

\bibitem{xia} D.~X.~Xia, {\em Measure and Integration Theory on
Infinite-Dimensional Spaces} (New York: Academic Press 1972).

\end{enumerate}
\end{document}